\documentclass[prd, preprint, aps, showpacs]{revtex4}

\usepackage{graphicx}
\usepackage{color}

\def\be{\begin{equation}}
\def\ee{\end{equation}}
\def\ba{\begin{eqnarray}}
\def\ea{\end{eqnarray}}
\def\beq{\begin{eqnarray}}
\def\eeq{\end{eqnarray}  }

\def\eref#1{Eq.~(\ref{#1})}

\def\fref#1{Fig.~\ref{#1}}
\def\Fref#1{Fig.~\ref{#1}}
\def\sref#1{Section~\ref{#1}}

\def\tref#1{Table~\ref{#1}}

\def\rmd{{\rm d}}
\def\rmO{{\rm O}}
\def\ADM{{\rm ADM}}
\def\HOR{{\rm H}}
\def\ISCO{{\rm ISCO}}
\def\BE{{E_b}}
\def\lp{\left(}
\def\rp{\right)}

\def\etal{{et al.}}

\begin{document}
\title{Physics and Initial Data for Multiple Black Hole Spacetimes}

\author{Erin Bonning}
\author{Pedro Marronetti}
\altaffiliation{Department of Physics, University of Illinois at
Urbana-Champaign}
\author{David Neilsen}
\altaffiliation{Department of Physics \& Astronomy, Louisiana State University,
Baton Rouge LA 70803}
\author{Richard Matzner}
\affiliation{Center for Relativity, University of Texas at Austin,
Austin, TX 78712-1081, USA}

\begin{abstract} An orbiting black hole binary will generate
strong gravitational radiation signatures, making these binaries important
candidates for detection in gravitational wave observatories.
The gravitational radiation is characterized by the orbital
parameters, including the frequency and separation at the inner-most
stable circular orbit (ISCO). One
approach to estimating these parameters relies on a sequence of initial
data slices that attempt to capture the physics of the inspiral.  Using
calculations of the binding energy, several authors have estimated the
ISCO parameters using initial data constructed with various
algorithms.  In this paper we examine this problem using conformally
Kerr-Schild initial data. We present convergence results for our
initial data solutions, and give data from numerical solutions of the
constraint equations representing a range of physical configurations.
In a first attempt to understand the physical content of the initial
data, we find that the Newtonian binding energy is contained in the
superposed Kerr-Schild background before the constraints are solved.
We examine some deficiencies with the initial data approach to orbiting
binaries, especially touching on the effects of prior motion and
spin-orbital coupling of the angular momenta.  Making rough
estimates of these effects, we find that they are not insignificant
compared to the binding energy, leaving some doubt of the utility of
using initial data to predict ISCO parameters. In computations of
specific initial-data configurations we find spin-specific effects
that are consistent with analytical estimates.

\end{abstract}


\maketitle

%----------------------------------------------------------------------
%
%
%
%----------------------------------------------------------------------
\section{Introduction}
\label{sec:intro}

The computation of gravitational wave production from the interaction
and merger of compact astrophysical objects is a 
challenge which, when solved, will provide a predictive and analytical
resource for the upcoming gravitational wave detectors. A binary black
hole system is expected to be the strongest possible astrophysical
gravitational wave source. In particular, one expects a binary black
hole system to progress through a series of quasi-equilibrium states of
narrowing circular orbits as it emits gravitational radiation.  In the
final moments of stellar mass black hole inspiral, the
radiation will be detectable in the current (LIGO-class) detectors. If
the total binary mass is of the order of $10M_{\odot}$, the moment of
final plunge to coalescence will emit a signal detectable by the
current generation of detectors from very distant (Gpc) sources.

Detecting gravitational radiation is also a significant technical
challenge.  Gravitational waves couple very weakly to matter, and the
expected signals are much smaller in amplitude than ambient environmental
and thermal noise.  The successful detection of these waves, therefore,
requires some knowledge of what to look for.  
In this regard, an orbiting binary black hole system is an ideal candidate
for detection since the orbital motion produces regular gravitational 
radiation patterns. In such an inspiraling
black hole system, the strongest waves are emitted during the last
several orbits, as the holes reach the innermost quasi-stable orbit
(here abbreviated ISCO), and as they continue through the final
plunge.  The dynamics of the holes during these final orbits, especially
the orbital angular velocity, $\omega_\ISCO$, and separation, $\ell_\ISCO$,
determine the dominant characteristics of the detectable waves.
Any knowledge of these parameters is advantageous for detecting
radiation from these binary systems.  

The proper way to predict gravitational waveforms for orbiting black
holes is to set initial data for two widely separated holes, and then
solve the evolution equations to follow the inspiral through merger and
beyond.  This problem is well beyond the capabilities of current
evolution codes.  Therefore, to obtain some information about orbiting black
holes we, and 
others~\cite{Cook,Pfeiffer,Baumgarte,GGB1,GGB2,GGB3,Cook2,Pfeiffer2}, 
turn to the initial value 
problem.  For an introduction to the literature, see the review by
Cook~\cite{CookReview}.
Given a collection of initial data for black holes in circular
orbits with decreasing radius, one tries to identify a sequence of initial
data that corresponds to instantaneous images of a time-dependent
evolution. Circular orbits are chosen because orbits in the early
stages of an inspiral are predicted to become circularized
because of the stronger gravitational radiation near periapse~\cite{Peters}. 
When a suitable sequence of initial data slices has been obtained, they can
then be used to determine various orbital parameters.  For example, the
change in binding energy with respect to the orbital radius allows one
to identify $\ell_\ISCO$, and a similar analysis of the angular
momentum gives $\omega_{\rm ISCO}$.  The difficulty in this approach
comes in ensuring that the initial data at one radius
correspond to the same physical system as the data for another
radius.  This can be done for some systems by using 
conserved quantities. For example, in the case of neutron stars, constant 
baryon number is 
an unambiguous indicator of the sameness of the stars. However, in black hole 
physics it is not available; it is unclear how to determine that two black
hole initial data sets do, in fact, represent the same physical system.

The initial data approach to studying binary black holes is thus not
without problems. These difficulties fall in two broad areas.  First,
there is no unambiguous way to set initial data in general relativity.
The current algorithms all require some arbitrary
mathematical choices to find a solution. For instance, the approach we
take requires the definition of the topology of a background space and
of its metric and the momentum of the metric, followed by solution of four
coupled elliptic differential equations for variables that adjust the 
background fields. But the choice of the background quantities is 
arbitrary to a
large extent.  The physical meaning of these mathematical choices is
not completely clear, but the effect is unmistakable.  Data constructed
with various algorithms can differ substantially, even when attempting
to describe the same physical system~\cite{Pfeiffer2}.  The data sets can be
demonstrated in many circumstances to contain the expected Newtonian
binding energy, as we show below (i.e., the binding energy of order
$\rmO(m^2/\ell)$ agrees with the Newtonian result at this order).
However, the data can differ significantly at $\rmO(m (m/\ell)^2)$.
These differences are attributed to differences in wave content of the
data which may reflect possible prior motion or may simply be
spurious.  At present it is neither possible to build prior motion into
the initial data, nor to specify how radiation is added to the the
solution, nor to know how much there is.  It is known that the
circular orbits and the ISCO so determined are in fact
method-dependent. Furthermore, the methods need not even agree that a
specific dataset represents a circular orbit; their subsequent
evolutions may not agree~\cite{Will}.

A second problem---and the principal physical difficulty with the initial 
data method for studying black
hole binaries---is the lack of unambiguous conserved quantities.
The best candidate for an invariant quantity is the event horizon area, 
$A_{\rm H}$.
This area is unchanging for isentropic processes due to the proportionality
of $A_{\rm H}$ with the black hole entropy. 
One can argue that since the quasi-circular orbit is quasi-adiabatic,
$A_{\rm H}$  is nearly invariant over some phase of the inspiral.  
But the inspiral cannot be completely adiabatic because it cannot
be made arbitrarily slow; the black holes will absorb an 
unknown amount of gravitational radiation while in orbit and will thereby 
increase in size.
Moreover, the event horizon is a global construct of the spacetime,
and cannot be determined from a single slice of initial data.  
Therefore, one must use the apparent horizon area, $A_{\rm AH}$,
as an {\it ersatz} invariant for initial data 
studies~\cite{Shoemaker,Huq}.
When the hole is approximately stationary,
these horizon areas may be nearly equal~\cite{Dreyer}.
In dynamic configurations---as should be appropriate for orbiting 
holes---these horizon areas may differ 
substantially~\cite{Caveny,Caveny1}.

We will investigate physical content of initial data,
focusing on Kerr-Schild spacetimes.  We examine binding energy to leading 
order, and find that in our method of constructing 
the superposed Kerr-Schild data, the background fields contain the
Newtonian binding energy: the subsequent solution of the elliptic equations
yields a only small correction.  Using numerical solutions we present
orbital configurations with solved initial data. We give a qualitative
discussion of physical effects that may confound any attempt to study
inspiral via a sequence of initial data, and which may affect the
determination of the location of the ISCO. We give some computational
examples consistent with these qualitative predictions.

%----------------------------------------------------------------------
%
%
%
%----------------------------------------------------------------------
\section{Review of initial data construction in general relativity}

In the computational approach we take a Cauchy formulation
(3+1) of the ADM type, after Arnowitt, Deser, and
Misner~\cite{ADM}. In such a method the 3-metric $g_{ij}$ is the fundamental
variable. The 3-metric and its momentum are specified at one initial time on
a spacelike hypersurface.  The ADM metric is  
\be
\rmd s^2 = -(\alpha^2 - \beta_i \beta^i)\,\rmd t^2 + 2\beta_i \, \rmd t \,\rmd x^i
     + g_{ij}\, \rmd x^i\, \rmd x^j
\label{eq:admMetric}
\ee
where $\alpha$ is the lapse function and $\beta^i$ is the shift 3-vector.
Latin indices run $1,2,3$ and are lowered and raised by $ g_{ij}$
and its 3-dimensional inverse $ g^{ij}$. $\alpha$ and $\beta^i$ are gauge
functions that relate the coordinates on each hypersurface to each other.
The extrinsic curvature, $K_{ij}$, plays the role of momentum conjugate
to the metric, and describes the embedding of a $t=\hbox{\rm constant}$
hypersurface into the 4-geometry.

The Einstein field equations contain both hyperbolic evolution equations,
and elliptic constraint equations.  
The constraint equations for vacuum in the ADM decomposition are:
\beq
R - K_{ij}K^{ij} + K^2 &=& 0,
\label{eq:constraintH}
\eeq
\beq
\nabla_j \lp K^{ij} - g^{ij}K\rp  &=& 0.
\label{eq:constraintK}
\eeq

Here $R$ is the 3-dimensional Ricci scalar, and $\nabla_j$ is the
3-dimensional covariant derivative compatible with $ g_{ij}$.  These
constraint equations guarantee a kind of transversality of the
momentum (\eref{eq:constraintK}). Initial data must satisfy these constraint
equations; one may not freely specify all components of $g_{ij}$ and
$K_{ij}$.  The initial value problem in general relativity thus
requires one to consistently identify and separate constrained and
freely-specifiable parts of the initial data.  Methods for making
this separation, and solving the constraints as an elliptic system,
include:  the {\it conformal transverse-traceless
decomposition}~\cite{YP}; the {\it physical transverse-traceless
decomposition}~\cite{MY}:  and the {\it conformal thin sandwich
decomposition} which assumes a helical killing 
vector~\cite{Mathews, York,GGB1}. 
These methods all involve
arbitrary choices and do not produce equivalent data. Our solution
method uses the conformal transverse-traceless decomposition~\cite{YP}. 

Solutions of the initial value problem have been addressed in
the past by several groups~\cite{Cook,YP,Baumgarte,Pfeiffer,GGB1}. It is the
case that until recently, most data have been constructed assuming that
the initial 3-space is conformally flat. The method most commonly used
is the approach of Bowen and York~\cite{Bowen+York}, which chooses
maximal spatial hypersurfaces and takes the spatial 3-metric to be
conformally flat. This method has been used to find candidate
quasi-circular orbits by Cook~\cite{Cook}, Baumgarte~\cite{Baumgarte}, and
most recently, Pfeiffer {\it et al.}~\cite{Pfeiffer}. 

The chief advantage of the maximal spatial hypersurface approach is
numerical simplicity, as the choice $K = 0$ decouples the Hamiltonian
constraint from the momentum constraint equations.  If, besides  $K =
0$, the conformal background is flat Euclidean 3-space, there are known
$K_{ij}$ that analytically solve the momentum
constraint~\cite{Bowen+York}. The constraints then reduce to one
elliptic equation for the conformal factor $\phi$.  However, it has
been pointed out by Garat and Price~\cite{Garat} that there are no
conformally flat $K=0$ slices of the Kerr spacetime. Since we expect
astrophysical sources to be rotating, the choice of a conformally flat
$K=0$ background will yield data that necessarily contains some
quantity of ``junk"  gravitational radiation. Jansen {\it et al.}~\cite{Jansen}
 have recently shown by comparison with known solutions that conformally flat
data do indeed contain a significant amount of unphysical gravitational
field.  Another conformally flat $K=0$ method recently 
used by Gourgoulhon, Grandclement, and
Bonazzola~\cite{GGB1,GGB2} is a  thin sandwich approximation based on the
approach of Wilson and Mathews~\cite{Mathews} which assumes the presence of an
instantaneous rotation Killing vector to define the initial extrinsic
curvature. They impose a specific gauge defined by demanding that $K$ and the 
conformal factor remain constant in the rotating frame.
  Since $\phi$ and $K$ are a conjugate pair 
in the ADM approach, this method solves the four initial value equations and 
one second-order evolution equation. The assumption of a Killing
vector suppresses radiation or, perhaps more accurately, imposes a
condition of equal ingoing and outgoing radiation.

In this paper we use Kerr-Schild data~\cite{Matzner:1999pt} to outline some 
of the difficulties in
finding the ISCO using the initial data technique. We discuss the extent
to which initial data set by means of superposed Kerr-Schild black holes limits
the extraneous radiation in the data sets, and we estimate the accuracy
of the extant published ISCO determinations.
Recent works by Pfeiffer, Cook, and Teukolsky also investigate binary black
hole systems using Kerr-Schild initial data~\cite{Pfeiffer2}.

%----------------------------------------------------------------------
%
%
%
%----------------------------------------------------------------------
\section{Initial Data \lowercase{via} Superposed Kerr-Schild Black Holes}
\label{sec:ks_id}

The superposed Kerr-Schild method for setting black hole initial data,
developed by Matzner, Huq, and Shoemaker~\cite{Matzner:1999pt},
produces data for black holes of arbitrary masses, boosts, and spins
without relying on any underlying symmetries of any particular
configuration.  The method proceeds in two parts.  First, a background
metric and background extrinsic curvature are constructed by
superposing individual Kerr-Schild black hole solutions.  Then the
physical data are generated by solving the four coupled constraint
equations for corrections to the background.  Intuitively, the
background solution should be very close to the genuine solution when
the black holes are widely separated, and only small adjustments to the
gravitational fields are required to solve the constraints. We show
that this is true for large and also for small separations. This
section briefly reviews the superposed Kerr-Schild method for initial
data, then gives some analytic results to justify this contention.

%----------------------------------------------------------------------
%
%
%
%----------------------------------------------------------------------
\subsection{Kerr-Schild data for isolated black holes} 
\label{sec:ks}

The Kerr-Schild~\cite{KerrSchild} form of a black hole solution describes the 
spacetime of a single black hole
with mass, $m$, and specific angular momentum, $a = j/m$, in a coordinate
system that is well behaved at the horizon.
(We use uppercase $M$ for calculated masses, e.g., the ADM mass,
and lowercase $m$ for mass parameters, or when the distinction 
is not important.)
The Kerr-Schild metric is
\be
        \rmd s^{2} = \eta_{\mu \nu}\,\rmd x^{\mu}\, \rmd x^{\nu} 
                 + 2H(x^{\alpha}) l_{\mu} l_{\nu}\,\rmd x^{\mu}\,\rmd x^{\nu},
        \label{eq:1}
\ee
where $\eta_{\mu \nu}$ is the metric of flat space, $H$ is a scalar 
function of $x^\mu$, and $l_{\mu}$ is an (ingoing) null vector, null 
with respect to both the background and the full metric,
\be
\eta^{\mu \nu} l_{\mu} l_{\nu} = g^{\mu \nu} l_{\mu} l_{\nu} = 0.
\label{eq:2}
\ee
This last condition gives $l^0 l_0 = - l^i l_i$.

The general non-moving 
black hole metric in Kerr-Schild form (written in Kerr's original 
rectangular coordinates) has 
\begin{equation}
        H = \frac{mr}{r^{2} + a^{2}\cos^{2} \theta},
        \label{eq:ks_h}
\end{equation}
and
\begin{equation}
        l_{\mu} = \left(1, \frac{rx + ay}{r^{2} + a^{2}}, \frac{ry - 
        ax}{r^{2} + a^{2}}, \frac{z}{r}\right),
        \label{eq:4}
\end{equation}
where $r,~ \theta$ (and $\phi$) 
are auxiliary spheroidal coordinates,  $z = r(x,y,z) \cos \theta$, 
and $\phi$ is 
the axial angle.  $r(x, y, z)$ is obtained from the relation,
\begin{equation}
        \frac{x^{2} + y^{2}}{r^{2} + a^{2}} + \frac{z^{2}}{r^{2}} = 1,
        \label{eq:5}
\end{equation}
giving
\begin{equation}
        r^{2} = \frac{1}{2}(\rho^{2} - a^{2}) +
        \sqrt{\frac{1}{4}(\rho^{2} - a^{2})^{2} + a^{2}z^{2}}, 
        \label{eq:6}
\end{equation}
with
\begin{equation}
\rho = \sqrt{x^{2} + y^{2} + z^{2}}.
        \label{eq:rho_def}
\end{equation}

Comparing the Kerr-Schild metric with the ADM 
decomposition~\eref{eq:admMetric}, we find that the $t=\hbox{\rm constant}$ 
3-space metric is: 

\be
g_{ij} = \delta_{ij} + 2 H l_i l_j,
\label{eq:3metric_ks}
\ee

Further, the ADM gauge variables  are
\be
\beta_i = 2 H l_0 l_i,
\label{eq:beta_ks}
\ee
and
\be
\alpha = \frac{1}{\sqrt{1 + 2 H l_0^2}}.
\ee

The extrinsic curvature can be computed from the metric using the ADM 
evolution equation~\cite{MTW}
\be
        K_{ij} = \frac{1}{2\alpha}[\nabla_j\beta_{i} + \nabla_i\beta_{j} 
                     - \dot g_{ij}],
\label{eq:k_ks}
\ee
where a dot ( $\dot{}$ ) denotes a the partial derivative with respect
to time.
Each term on the right hand side of this equation is known analytically.

%----------------------------------------------------------------------
%
%
%
%----------------------------------------------------------------------
\subsection{Boosted Kerr-Schild black holes} 

The Kerr-Schild metric is form-invariant under a 
boost, making it an ideal metric to describe moving
black holes.  A constant Lorentz transformation 
(the boost velocity, ${\bf v}$, is specified with respect to the background
Minkowski spacetime) $\Lambda^{\alpha}{}_{\beta}$ leaves the
4-metric in Kerr-Schild form, with $H$ and $l_{\mu}$
transformed in the usual manner:\\
\ba
  x'^{\beta} &=& \Lambda^\beta{}_\alpha x^{\alpha},\\ 
  H'(x'^{\alpha})  &=&  H\lp (\Lambda^{-1})^\alpha{}_\beta  
                          \,\,x'^{\beta}\rp,\\ 
  l'_{\delta}(x'^{\alpha}) &=& \Lambda^{\gamma}{}_{\delta}\,\, 
            l_{\gamma}\lp(\Lambda^{-1})^\alpha{}_\beta\,\, x'^{\beta}\rp .
\label{eq:ks_boost}
\ea
Note that $l'_{0}$ is no longer unity. As the initial solution
is stationary, the only time dependence comes in the
motion of the center, and the full metric is stationary with a Killing
vector reflecting the boost velocity.
The solution, therefore, contains no junk radiation, as no radiation 
escapes to infinity during a subsequent evolution.
Thus, Kerr-Schild data exactly represent a spinning and/or moving single
black hole. This is not possible in some other approaches, e.g.,
the conformally flat approach~\cite{Jansen}.

%----------------------------------------------------------------------
%
%
%
%----------------------------------------------------------------------
\subsection{Background data for multiple black holes}

The structure of the Kerr-Schild metric suggests a natural extension
for multiple black hole spacetimes using the straightforward superposition
of flat space and black hole functions
\be
{g}_{ij} \approx \eta_{ij}
            + 2~{}_1H_{~1}l_{i~1}l_{j}
            + 2~{}_2H_{~2}l_{i~2}l_{j}
            + \cdots\, ,
\label{eq:metric_super_simple}
\ee
where the preceding subscript numbers the black holes.  Note that a
simple superposition is typically {\em not} a genuine solution of the Einstein
equations, as it does not
satisfy the constraints, but it should be ``close'' to the real solution
when the holes are widely separated.  

To generate the background data,
we first choose mass and angular momentum parameters for each hole,
and compute the respective $H$ and $l^\alpha$ in the appropriate 
rest frame.  These quantities are then boosted in the desired direction 
and offset to the chosen position in the computational frame.  
The computational grid is the center of momentum frame for the two holes, 
making the velocity of the second hole a function of the two
masses and the velocity of the first hole.
Finally, we compute the 
individual metrics and extrinsic curvatures in the coordinate system 
of the computational domain: 
\beq
   {}_A g_{ij} &=&  \eta_{ij} 
                     + 2~{}_A H ~{}_A l_{i} ~{}_A l_{j},\\
   {}_A K_i{}^m &=& \frac{1}{2\alpha} ~{}_A g^{mj}
               \lp \nabla_j ~{}_A\beta_{i} + \nabla_i~{}_A \beta_{j} 
                    - ~{}_A \dot g_{ij}\rp.
\eeq
Again, the index $A$ labels the black holes.
Data for $N$ holes are then constructed in superposition
%\beq
%\tilde{g}_{ij} &=& \eta_{ij} + 2~{}_{1}B{}_{~1}H_{~1}l_{i~1}l_{j}
%            + 2~{}_{2}B{}_{~2}H_{~2}l_{i~2}l_{j}\,,\nonumber\\
%\tilde{K} &=& {}_{1}B{}_{~1}K_i^{~i}+{}_{2}B{}_{~2}K_i^{~i}\,,\\
%\tilde{A}_{ij} &=& \tilde{g}_{n(i}~~({}_{1}B{}_{~1}K_{j)}^{~n}
%        +{}_{2}B{}_{~2}K_{j)}^{~n}
%        - \frac{1}{3} \delta_{j)}^{~n} \tilde{K})\, .\nonumber
%\label{eq:ks_super}
%\eeq
%

\beq
\tilde{g}_{ij} &=& \eta_{ij} + \sum_A^N 2~{}_AB~{}_A H {}_A l_i ~{}_A l_j ,\\
\tilde{K} &=& \sum_A^N {}_AB~{}_AK_i{}^i ,\\
%\tilde{A}_{ij} &=& \tilde{g}_{n(i}~~\sum_A^N \lp {}_AB~{}_AK_{j)}{}^n
%                  -\frac{1}{3} \delta_{j)}{}^n {}_AB~{}_AK_i{}^i\rp .
\tilde{A}_{ij} &=& \tilde{g}_{n(i}~~\sum_A^N {}_AB~\lp {}_AK_{j)}{}^n
                  -\frac{1}{3} \delta_{j)}{}^n ~{}_AK_i{}^i\rp .
\label{eq:ks_super}
\eeq
A tilde ( $\tilde{}$ ) indicates a background field tensor.
The simple superposition of the metric from~\eref{eq:metric_super_simple} 
(part of the original specification~\cite{Matzner:1999pt}) has been modified
here with the introduction of 
{\it attenuation functions}, ${}_A B$~\cite{Marronetti,Marronetti:2000rw}. 
The extrinsic curvature is separated into its trace, $K$, and trace-free
parts, $A_{ij}$, and the indices of $\tilde{A}_{ij}$ are 
explicitly symmeterized.  

The attenuation functions represent the physical idea that in the immediate
vicinity of one hole, the effect of a second hole becomes negligible.
Near a black hole the conformal background
superposition ( $ \tilde{} $ ) metrics approach the analytic values for
the single black hole.
The attenuation
function ${}_{2}B$ (${}_1B$) eliminates the
influence of the second (first) black hole in the vicinity of the first
(second).  
${}_{1}B$ equals unity everywhere except in the vicinity of the second
black hole, and its first and second derivatives are zero at the
singularity of the second hole. 

\begin{figure}
\begin{center}
\includegraphics[width=5.in]{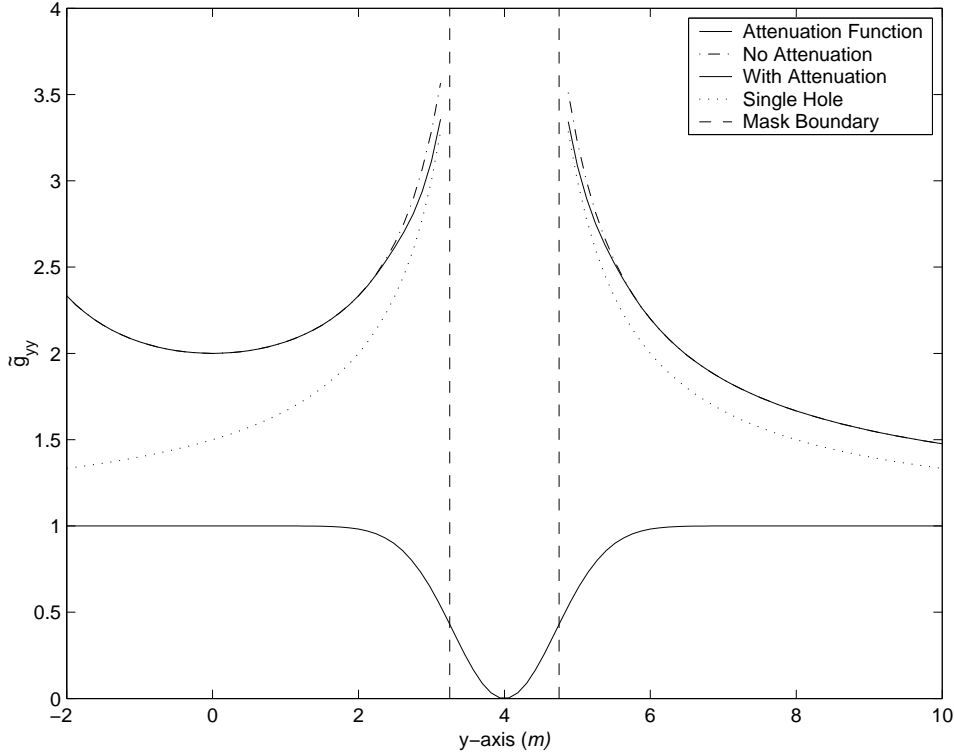}
\end{center}
\caption{
The attenuation function,
${}_1 B =1-\exp(-\ell_{1}^{4}/2\sigma^2)$, used to calculate our initial
data solutions.  
To indicate the effect of the attenuation function in a binary black 
hole system, we also plot the 
the background metric function $\tilde g_{yy}$
in the vicinity of one hole with and without attenuation.  
The Schwarzschild black holes are placed along the $y$-axis at $\pm 4 m$.
Here  $\ell_1$ is the coordinate distance from the center 
of the second black hole, and the attenuation function width is  
$\sigma = m^2$.
} 
\label{fig:attenuation}
\end{figure}

The attenuation function used is

\be
{}_{1}B =1-\exp(-\ell_{1}^{4}/2\sigma^2),
\ee
where $\ell_1$ is the
coordinate distance from the center of hole $2$, 

\ba
\ell_1^{2} &=& \frac{1}{2}(\rho^{2} - a^{2}) +
\sqrt{\frac{1}{4}(\rho^{2} - a^{2})^2 + a^{2}z^{2}}~~,\\
\rho &=& \sqrt{{}_{2}\gamma^2(x-{}_{2}x)^{2} + (y-{}_{2}y)^{2}
+ (z-{}_{2}z)^{2}}~.
\ea

\noindent and $\sigma$ is a parameter.  In all examples given in this paper, 
the masses are equal and $\sigma = m^2$.
\Fref{fig:attenuation} shows a typical attenuation function used in 
calculating our initial data sets. 

A small volume containing the singularity
is masked from the computational domain.
This volume is
specified by choosing a threshold value for the Ricci scalar, typically 
for $|R| \geq 2/m^2$. For a single Schwarzschild black hole, this gives a
spherical mask with a radius $r \simeq 0.73~m$. In all cases the masked
region lies well within apparent horizons in the solved data.  In
practice we find that a small attenuation region (also inside the
apparent horizon) is necessary to achieve a smooth solution of the
elliptic initial data equations near the mask; see Section {\bf II.D} below.  
Figures \ref{fig:hc_with_att} and
\ref{fig:mc_with_att} show the Hamiltonian and momentum constraints for
the background space with and without attenuation. We have not varied
the masking condition to determine what effect the size of the mask has
on the global solution.  As mentioned below, Pfeiffer {\it et al.}
have investigated this point~\cite{Pfeiffer2}.

\begin{figure}
\begin{center}
\includegraphics[width=3.5in,angle=270]{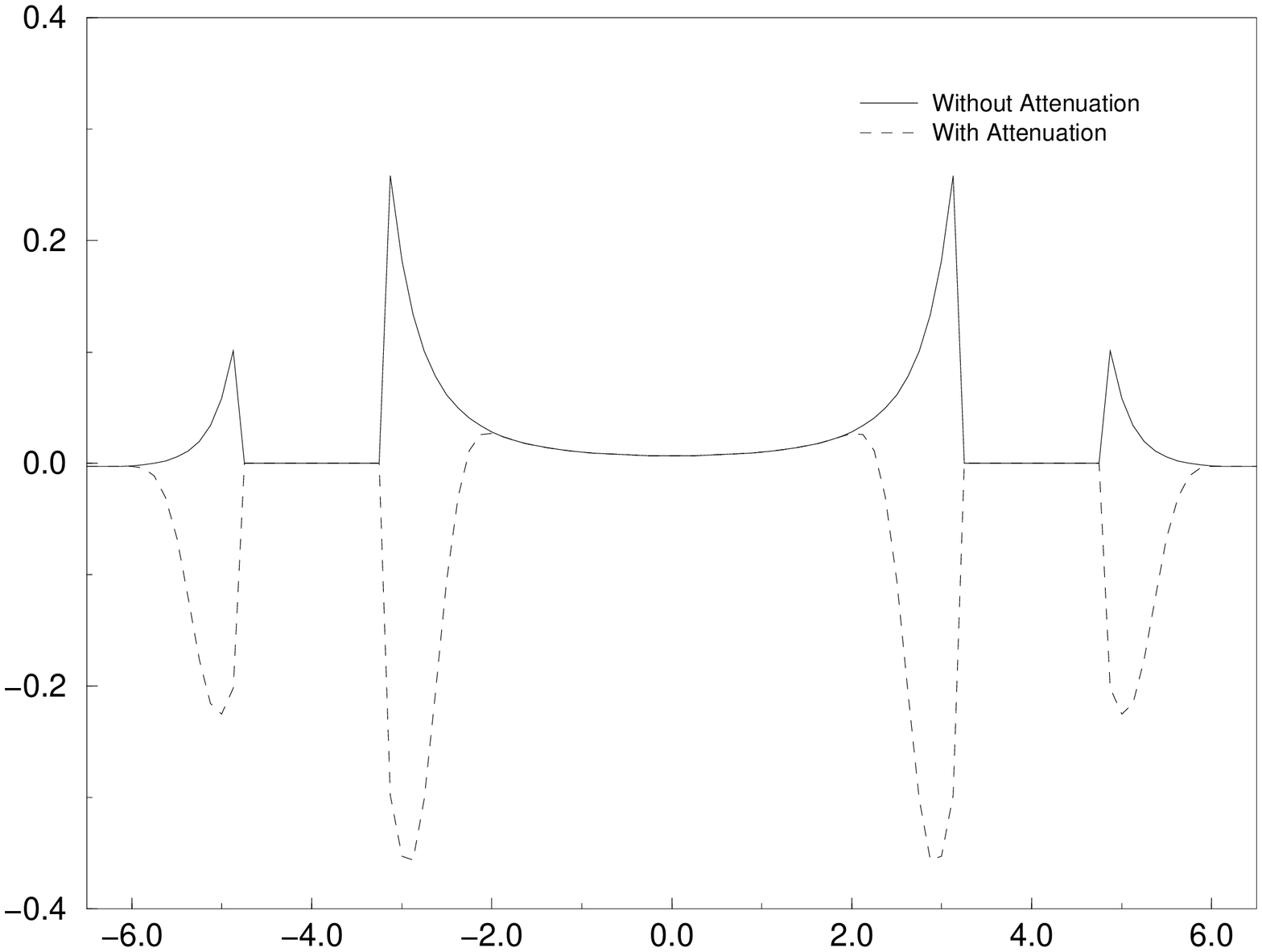}
\end{center}
\caption{The Hamiltonian constraint (units $m^{-2}$) calculated for the
background space for two identical Schwarzschild black holes.  
The black holes are located on the $y$-axis at $y = \pm 4$~m, 
and have zero initial velocity.
The solid curve is the background behavior of the constraint
without using attenuation functions, and the dashed curve is the
constraint
with attenuation and $\sigma = m^2$. The masked region is within the
radius $r ~~\widetilde{<}~~0.73 m$. It can be seen that attenuation 
does not necessarily reduce the constraint, but does smooth it.}
\label{fig:hc_with_att}
\end{figure}

\begin{figure}
\begin{center}
\includegraphics[width=3.5in,angle=270]{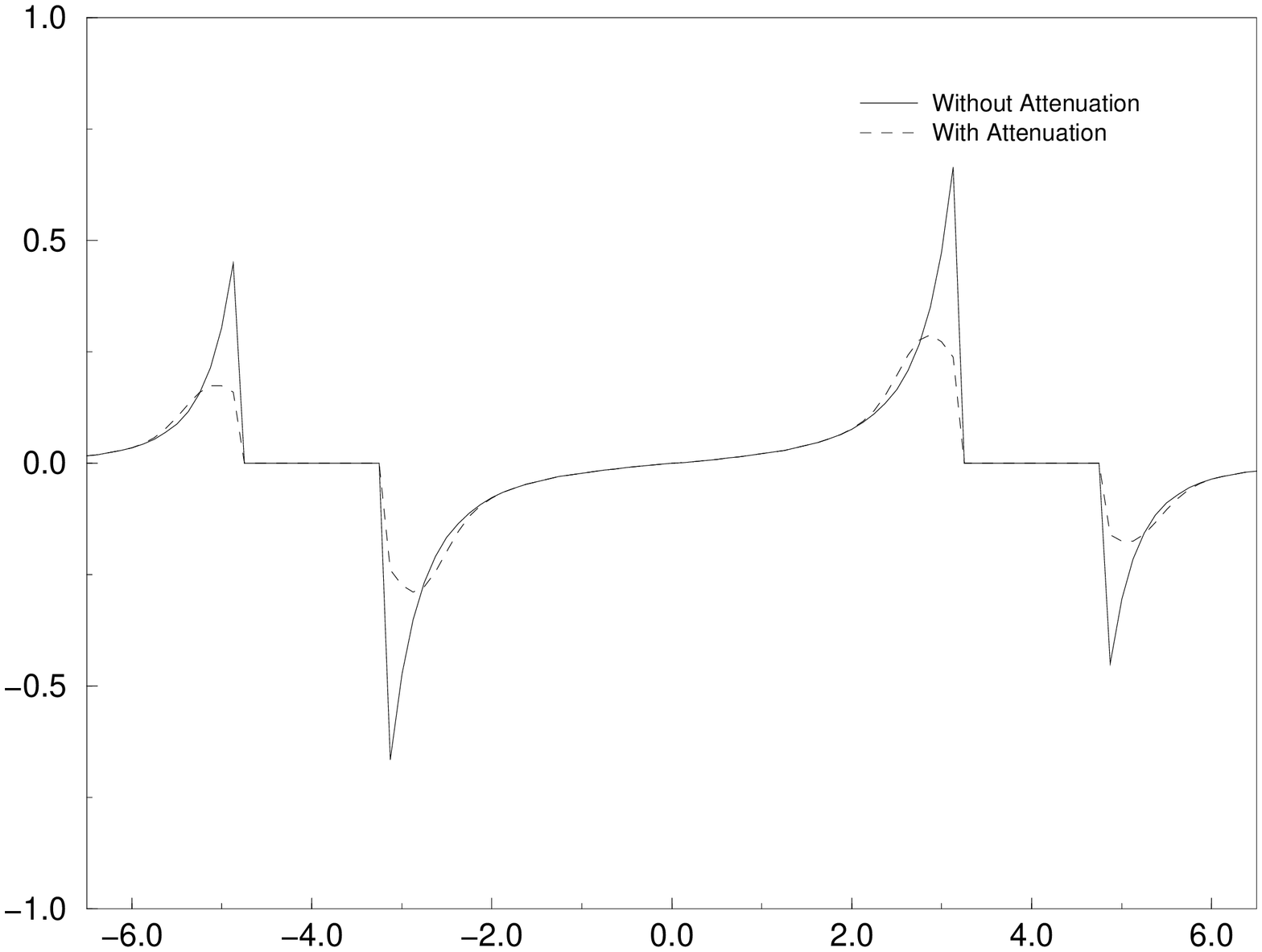}
\end{center}
\caption{The $y$-component of the momentum constraint (units $m^{-2}$)
calculated for the background space of two identical Schwarzschild 
black holes.  The black holes are located on the $y$-axis at $y = \pm 4~m$,
and have zero initial velocity.  The solid curve is the background
behavior
of the constraint
without using attenuation functions, and the dashed curve is the
constraint
with attenuation and $\sigma = m^2$.
}
\label{fig:mc_with_att}
\end{figure}

%----------------------------------------------------------------------
%
%
%
%----------------------------------------------------------------------
\subsection{Generating the physical spacetime}

The superposition of Kerr-Schild data described in the previous section
does not satisfy the constraints,
Eqs.~(\ref{eq:constraintH})--(\ref{eq:constraintK}), and hence are not
physical. 
A physical spacetime can be constructed by modifying the background
fields with new functions such that the constraints are satisfied.
We adopt the conformal transverse-traceless method
of York and collaborators~\cite{YP} which consists of a
conformal decomposition 
and a vector potential that adjusts the longitudinal components of the 
extrinsic curvature.
The constraint equations are then solved for these new quantities such that
the complete solution fully satisfies the constraints.

The physical metric, $g_{ij}$, and the trace-free part of the extrinsic 
curvature, $A_{ij}$, are related to the background fields through a conformal
factor
\ba
g_{ij} &=& \phi^{4} \tilde{g}_{ij}, \label{confg1} \\
\label{confg}
A^{ij} &=& \phi^{-10} (\tilde{A}^{ij} + \tilde{(lw)}^{ij}),
\label{eq:conf_field}
\ea
where $\phi$ is the conformal factor, and $\tilde{(lw)}^{ij}$
will be used to cancel any possible longitudinal contribution to the 
superposed background extrinsic curvature. 
$w^i$ is a vector potential, and
\ba
\tilde{(lw)}^{ij} \equiv \tilde{\nabla}^{i} w^{j} + \tilde{\nabla}^{j} w^{i} 
        - \frac{2}{3} \tilde{g}^{ij} \tilde{\nabla_{k}} w^{k}.
\label{lw}
\ea
The trace $K$ is taken to be a given function
\be
K = \tilde K.
\label{tk}
\ee
Writing the Hamiltonian and momentum constraint equations in terms of
the quantities in 
Eqs.~(\ref{confg1})--(\ref{tk}), we obtain four coupled 
elliptic equations for the fields $\phi$ and $w^i$~\cite{YP}:
\ba
\tilde{\nabla}^2 \phi &=&  (1/8) \big( \tilde{R}\phi 
        + \frac{2}{3} \tilde{K}^{2}\phi^{5} -   \nonumber \\
        & & \phi^{-7} (\tilde{A}{^{ij}} + (\tilde{lw})^{ij})  
            (\tilde{A}_{ij} + (\tilde{lw})_{ij}) \big),   \\ 
\tilde{\nabla}_{j}(\tilde{lw})^{ij} &=& \frac{2}{3} \tilde{g}^{ij} \phi^{6} 
        \tilde{\nabla}_{j} K - \tilde{\nabla}_{j} \tilde{A}{^{ij}}.
\label{ell_eqs}
\ea

\subsection{Boundary Conditions}
\label{sec:boundary}

A solution of the elliptic constraint equations requires that boundary
data be specified on both the outer boundary {\em and } the surfaces
of the masked regions. This contrasts with the hyperbolic evolution 
equations for which excision can in principle be carried out without 
setting inner boundary data since no information can propagate 
out of the holes.  Boundaries in an elliptic system, on the other hand, 
have an immediate influence on the entire solution domain.   
Using the attenuation functions, we can choose simple 
conditions, $\phi = 1$ and $w^{i} = 0$, on the masked regions surrounding
the singularities.
In practice this inner boundary condition is not completely satisfactory
because it generates small discontinuities in the solution at this boundary.
These discontinuities are small relative to the scales in the problem,
and are contained within the horizon.
We have made no attempt to determine their global effect on the solution.
Pfeiffer \etal~\cite{Pfeiffer2} report a similar observation,
and note that the location of the boundary does affect some 
aspects of the solution, though it has little effect on the fractional binding 
energy or the location of the ISCO.

The outer boundary conditions are more interesting.  Several physical
quantities of interest, e.g., the ADM mass and momenta, 
are global properties of the spacetime, and are calculated 
on surfaces near the outer boundary of the computational
grid.   Hence the outer boundary conditions must be chosen carefully
to obtain the proper physics.  
We base our outer boundary conditions on an asymptotic expansion of the
Kerr-Schild metric, which relies on the ADM
mass and momentum formul\ae\ to identify the physically relevant terms
at the boundaries.  We first review these expansions and
formul\ae.

An asymptotic expansion of the Kerr-Schild metric ($\rho \gg m$)
gives
\beq
 r &=& \rho \left(1 + \rmO (\rho^{-2})\right),\\
 H &=& m/\rho \left( 1 + \rmO (\rho^{-2}) \right),\\
   \label{eq:h_inf}
 l_{i} &=& n_{i} + \frac{a^{c}\epsilon_{ijc} n^{j}}{\rho} + \rmO (
  \rho^{-2} ),
   \label{eq:l_inf}
\eeq
where 
$n_{i} = n^{i} = x^{i}/\rho$. 
(This is the only place where we do {\it not} use the 3-metric to raise 
and lower indices, and $n_i n^i = 1$).
$a^{c}$ is the Kerr spin parameter with a general direction:
$a^{c} = a \hat a^{c}$.
The shift (Eq.~\ref{eq:beta_ks}) is asymptotically
\begin{equation}
   \beta_{i} = \frac{2m}{\rho} \left( n_{i} + a^{c}
   \frac{\epsilon_{ijc}n^{j}}{\rho} \right) + \rmO(\rho^{- 3}).
   \label{eq:beta_inf}
\end{equation}
The asymptotic expansion of the extrinsic curvature in the stationary
Kerr-Schild form (cf.\ \eref{eq:k_ks}) is 
\begin{eqnarray}
   \alpha K_{ab} &=& 
       \frac{2m}{\rho^{2}} \left( -2n_{a}n_{b} + \delta_{ab}\right) \nonumber\\
   & & \hspace{0.1cm} - \frac{3m}{\rho^{3}} a^{c} 
                 \left( \epsilon_{ajc}n_{b} + \epsilon_{bjc}n_{a}\right)n_{j}
                         \nonumber\\
   & & \hspace{0.1cm}  
     + \frac{6m^2}{\rho^{3}} \left(n_{a}n_{b} -\frac{2}{3} \delta_{ab}\right) 
     + \rmO(\rho^{-4}) .
   \label{eq:k_inf}
\end{eqnarray}
The terms proportional to $a^c/\rho^3$ in this expression arise from the
transverse components of $\beta^a$ ($\beta_a n^a = 0$); the terms 
of $ \rmO(\rho^{-3})$ independent of 
$a^c$ arise from the affine connection. 
Note that $\alpha= 1+  \rmO(\rho^{-1})$, and will
not affect the ADM estimates below.

The ADM formul\ae\ are evaluated in an asymptotically flat region 
surrounding the system of interest, and in Cartesian coordinates they are
\beq
  \label{eq:adm_mass}
M_{\ADM} &=& \frac{1}{16\pi} \oint \left( \frac{\partial g_{ji}}{\partial
  x^{j}} - \frac{\partial g_{jj}}{\partial x^{i}} \right) 
  \rmd S^i,\\
   \label{eq:adm_mom}
P^{\ADM}_{k} &=& \frac{1}{8\pi} \oint \left( K_{ki} - K^{b}{}_{b} \delta_{ki}
   \right)\rmd S^i,\\
   \label{eq:adm_ang_mom}
J^{\ADM}_{ab} &=& \frac{1}{8\pi} \oint \left( x_{a}K_{bi} - x_{b}K_{ai}
   \right) \rmd S^i,
\eeq
for the mass, linear momentum, and angular momentum of the system
respectively~\cite{Wald, note4}.  (All repeated indices are summed.)
The mass and linear momentum together constitute a 4-vector under Lorentz
transformations in the asymptotic Minkowski space, and
the angular momentum depends only on the trace-free 
components of the extrinsic curvature.

To compute the ADM mass and moment\ae for a single, stationary Kerr-Schild 
black hole, we evaluate the integrals on the surface of a distant sphere.
The surface element then becomes
$dS^i = n^i \rho^2 \rmd\Omega$, where $n^i$ is the outward normal and
$\rmd\Omega$ is the differential solid angle.
we need to evaluate the metric only to order $\rmO(\rho^{-1})$;
the differentiation in Eq.~(\ref{eq:adm_mass}) guarantees that
terms falling off faster than $\rho^{-1}$ do not contribute to the
integration.  The integrand is then 
$ \frac{4m}{\rho^{2}} n_{i} \rho^{2}n^i \,\rmd\Omega$ 
and the integration yields the expected ADM mass $M_{ADM}=m$.
The ADM linear momentum requires only the leading order of of $K_{ab}$, 
$\rmO(\rho^{-2})$;
terms falling off faster than this do not contribute.  
The integrand of \eref{eq:adm_mom} then becomes
$-\frac{4m}{\rho^{2}} n_{a}n_{b} n^{b} \rho^{2} \,\rmd\Omega,$
yielding zero for the 3-momentum, as expected for a non-moving black
hole.

At first blush, the integral for the ADM angular 
momentum~\eref{eq:adm_ang_mom} appears
warrant some concern:
To leading order $K_{ab}$ is $\rmO({\rho^{-2}})$,
and the explicit appearance of $x_{a}$ in the integrand suggests that it
grows at infinity as $\rmO(\rho)$,
leading to a divergent result.
However, inserting the leading order term of $K_{ab}$ for a single, 
stationary Kerr-Schild black hole into the integrand of
\eref{eq:adm_ang_mom}, we find that the integrand is identically zero.
The $\rmO({\rho^{-2}})$ terms of $K_{ab}$ contain the quantities
$n_a n_b$ and $\delta_{ab}$, which separately cancel because of the 
antisymmetric form of \eref{eq:adm_ang_mom},
and a divergent angular momentum is avoided.
Including the $\rmO(\rho^{-3})$ terms of $K_{ab}$, we find
$J^\ADM_{ab} = \epsilon_{abc}a^{c}m$; the symmetry of the other
$\rmO(m^2\rho^{-3})$ terms again means they do not contribute.
This result for $J^\ADM_{ab}$ thus depends on
terms in the integrand proportional to $a$ that arise from corresponding 
terms in  $\beta^i$ proportional to $q^i$ where 
$q^a$ is a unit vector transverse to the radial direction, $q^a n_a=0$.
Only these terms contribute to
the angular momentum integral; in particular those terms in $\beta^i$ 
proportional to 
$n^i/\rho$ do not contribute. 

The ADM mass and momenta are Lorentz invariant. For a single, 
boosted black hole, we naturally obtain $M_\ADM = \gamma m$ and
$P_\ADM = \gamma m v$.
The background spacetime for multiple black holes is constructed 
with a superposition principle, and the ADM quantities are linear in 
deviations about flat space at infinity.  
Thus the ADM formul\ae, evaluated at infinity in the superposition, 
{\it do} yield the expected superposition.  For example, given two
widely separated
black holes boosted in the $x$-$y$ plane with spins aligned along
the $z$-axis, we have
\beq
\tilde M_\ADM &=&{}_1 \gamma {}_1 m + {}_2\gamma {}_2 m,\\
\tilde P^\ADM_i &=& 0, \\
\tilde J^\ADM_{12} &=& {}_1\gamma \left({}_1m {}_1v {}_1b  
                  + {}_1m {}_1a\right) + {}_2\gamma \left({}_2m {}_2v {}_2b  
                  + {}_2m {}_2a\right),
\label{eq:boosted}
\eeq
where ${}_1b$ and ${}_2b$ are impact parameters~\cite{note1},
and the tilde ( $ \tilde{} $ ) superscript
indicates that these quantities are calculated with the background tensors
$\tilde g_{ab}$ and $\tilde K_{ab}$.  This superposition principle for
the ADM quantities in the background data is one advantage of conformal
Kerr-Schild initial data.  (Note, in choosing the center of
momentum frame for the computation, $P^{\ADM}_{i} =0$ is a condition for
setting the background data.)

Consider now the ADM integrals for the solved data.  The Hamiltonian
constraint becomes an equation for the conformal factor, $\phi$.
As this equation is a nonlinear generalization of Poisson's equation,
asymptotic flatness in the full, solved metric requires that
\be
\phi \longrightarrow 1 + \frac{C}{2\rho} + \rmO(\rho^{-2}),
\ee
where $C$ is a (finite) constant.  
This leads to our outer boundary condition for $\phi$, namely
\begin{equation}
 \partial_{\rho} \left( \rho (\phi - 1) \right)|_{\rho \rightarrow
\infty} = 0.
\label{eq:phi_boundary}
\end{equation}
Furthermore, the linearity of the ADM mass integral gives
\begin{equation}
M_\ADM\mbox{(solved)} = {}_1\gamma{}_{1}m +
{}_2\gamma {}_{2}m + C.
\label{eq:m_adm_solved}
\end{equation}
(Here the absence of a tilde  ( $ \tilde{} $ )
indicates that this mass is calculated
using the solved $g_{ab}$.)  At this point we cannot predict even the
sign of $C$, though $|C|$ is expected to be small for widely separated
holes.  If $|C|\to \infty$, then the boundary condition 
\eref{eq:phi_boundary} would fail.  The existence of solutions using
this condition, however, provides evidence that this possibility does not 
occur.

The boundary condition for $w^{i}$ is more subtle.  {\it A priori}, we
expect $w^{i} \rightarrow 0$ at infinity, but a physically
correct solution on a finite domain requires that we understand how
$w^{i}$ approaches this limit at infinity.  
We construct our boundary conditions on $w^k$ by demanding that the 
ADM angular momentum of the full (solved) system be only finitely 
different from that of the background (superposed) data.
That is, given that $\{\tilde g_{ab}, \tilde K_{ab}\}$ and 
$\{g_{ab},K_{ab}\}$ have finite differences at infinity, we demand that
$J_{ab} -\tilde J_{ab}$ also be finite.
Using (\eref{eq:conf_field}) and (\eref{eq:adm_ang_mom}), 
we find for the difference in angular momentum
\be
   J_{ab}- \tilde J_{ab} = \frac{1}{8\pi} \oint\left( x_{a} 
\nabla_{(b} w_{i)} - x_{b}
   \nabla_{(a} w_{i)} \right)\rmd S^i\,.
\label{eq:j_adm_diff}
\ee
($\phi\to 1$ at infinity, and there is
no difference at this order between conformal and physical versions of 
$w^i$ and $g_{ab}$ 
at infinity.)

We have already evaluated an integral of this form, in the discussion
of the Kerr angular momentum (see \eref{eq:adm_ang_mom} and \eref{eq:k_inf}),
where we expressed $K_{ab}$ in terms of the Kerr-Schild shift vector.
In that analysis, we noted that falloff of the form
\be
   w_{i} \longrightarrow \frac{C_1}{\rho}n_{i} + \frac{C_2}{\rho^{2}}
   q_{i} + \rmO(\rho^{-3}),
\label{24}
\ee
with $C_1$ and $C_2$ constant, and $q_i n^i = 0$, 
will give a finite contribution to the angular momentum.
We therefore take as boundary conditions:
\begin{eqnarray}
   &  & \partial_\rho (\rho w^{i} n_{i}) = 0 \\[.12in]
\label{25}
   &  & \partial_\rho \left( \rho^{2} w^{i} (\delta_{ij} - n_{i}n_{j})
   \right) = 0\,.
\label{26}
\end{eqnarray}

\begin{figure}
\begin{center}
\includegraphics[width=3.0in, angle=270]{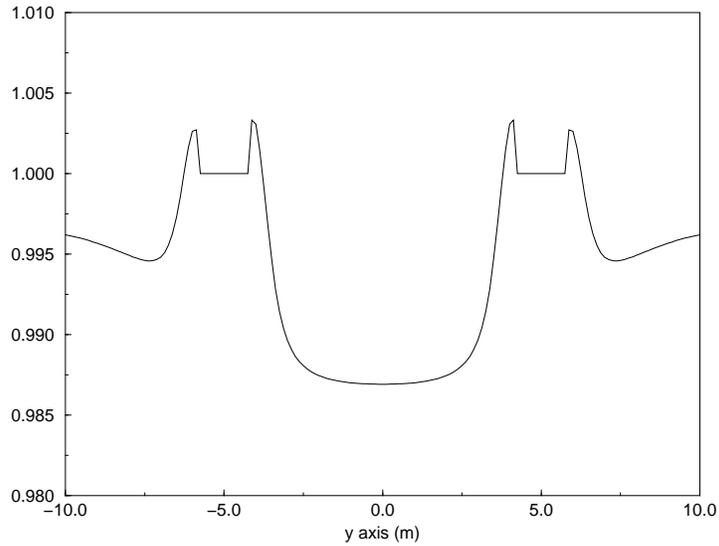}
\end{center}
\caption{$\phi$ along the y-axis connecting two nonspinning holes with 
orbital angular momentum. The holes are boosted in the $\pm$
x direction with $v=0.196$ and are separated by $10$~M. Note that $\phi$
is very close to unity everywhere.
}
\label{fig:circular_orbit_nospin_1phi}
\end{figure}

\begin{figure}
\begin{center}
\includegraphics[width=3.0in, angle=270]{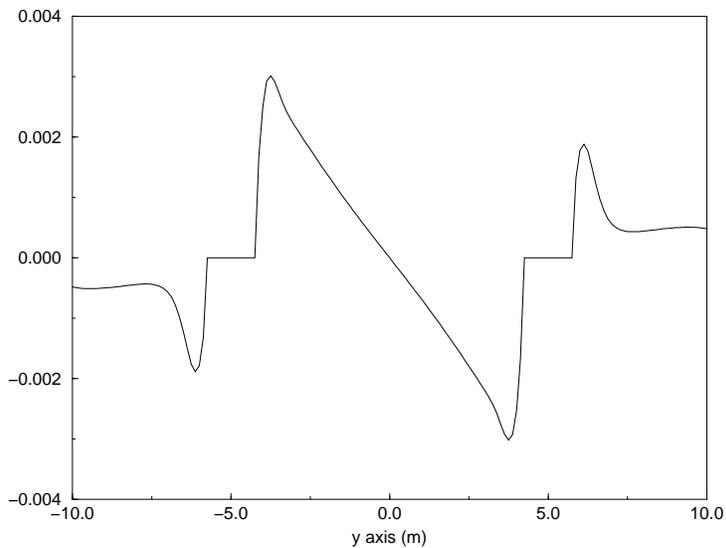}
\end{center}
\caption{$w^{x}$ for the same configuration as in Figure
\ref{fig:circular_orbit_nospin_1phi}. 
}
\label{fig:circular_orbit_nospin_1wx}
\end{figure}

\begin{figure}
\begin{center}
\includegraphics[width=3.0in, angle=270]{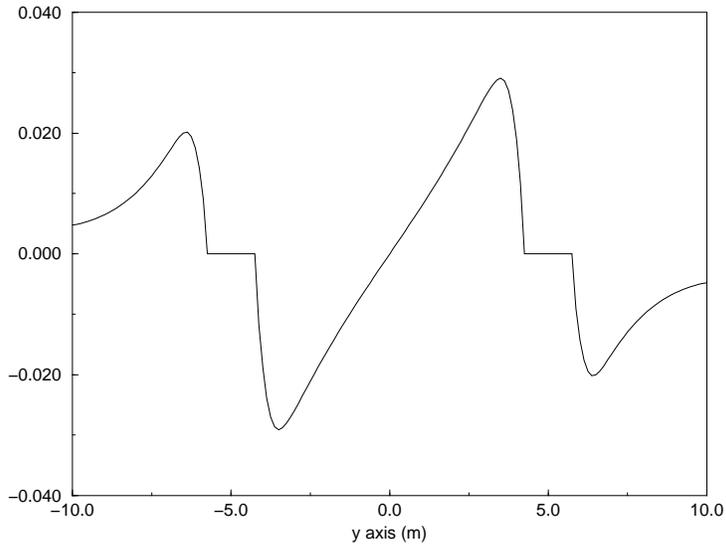}
\end{center}
\caption{$w^{y}$ for the same configuration as in
Figure \ref{fig:circular_orbit_nospin_1phi}.  $w^z$ is numerically
zero as expected by symmetry.
}
\label{fig:circular_orbit_nospin_1wy}
\end{figure}

Figures (\ref{fig:circular_orbit_nospin_1phi})--(\ref{fig:circular_orbit_nospin_1wy})
display $\phi$ and $w^i$ for a simple configuration. In this case the
elliptic equations were solved on a domain of $\pm 10~m$ along each axis
with resolution $\Delta x = m/8$.
As can be seen in these figures, the functions  $\phi$ and $w^i$ actually 
result in little adjustment to the background configurations. 
Also note that the radial component of $w^i$, $w^i n_i$,
is the dominant function.  In the graphs plotted here, which give the 
functions along the $y$-axis, we find 
of $||w^y||_\infty \approx 0.03$, while $||w^x||_\infty \approx 3
\times 10^{-3}$, and $||\phi -1||_\infty \approx 0.013$. 
Because of the symmetry of
the configuration, $||w^z||_\infty$ is much smaller. 
Analytically, $w^z = 0$ on the $y$-axis;
computationally we find $||w^z||_\infty \approx 5 \times 10^{-7}$. 
In fact we find in
general that the radial component of $w^i$ is the dominant function
in all directions, consistent with our boundary conditions, and  
consistent with the finding that solution of
the constraints has small effect on the computed angular momentum.
Of course the corrections $\phi$ and $w^i$ would be expected
to be larger, for data describing holes closer together. We show below that 
this data setting method leads to generically smaller corrections than
found in other methods, thus allowing closer control of the physical
content of the data.

%----------------------------------------------------------------------
%
%
%
%----------------------------------------------------------------------
\section{Binding Energy in Initial Data}
\label{sec:binding_energy}

As a first step towards understanding the physical content of initial
data sets,
we examine in this section the effect of the presence of a second hole
on the horizon areas of a first hole and on global features such
as the ADM integrals and the binding energy of the pair.  
This analysis is carried out for non-spinning holes to first order in the 
binding energy.
A comparison to the Newtonian result indicates that the 
Kerr-Schild {\it background} superposition data contain the appropriate
physical information at this level.   
We then consider possible spin-related phenomena, estimate
their magnitude, and discuss their possible effect near the ISCO.

%----------------------------------------------------------------------
%
%
%
%----------------------------------------------------------------------
\subsection{Binding energy in Brill-Lindquist data}
\label{sec:brill_lindquist}

Before discussing the conformal Kerr-Schild data, we first consider
Brill-Lindquist data for two non-moving Schwarzschild black holes~\cite{Brill}.  
These data are conformally flat, and $K_{ab} = 0$.  The momentum
constraints are trivially satisfied, and the Hamiltonian constraint is
solved for a conformal factor:  $\phi = 1 + m/(2r) + m'/(2r')$.
Here the two mass parameters are $m$, and $m'$, 
and $r$ and $r'$ are the distances in the flat background 
from the holes $m$ and $m'$.
  
We find that
the apparent horizon areas in the solved data correspond to 
\be
M_{\rm AH} + M_{\rm AH}' = m + m' + \frac{mm'}{\ell} +
\rmO(\ell^{-2}).
\ee
 The subscript 
``$\rm AH$''
indicates masses computed from apparent horizon areas, and the
separation in the flat background space is $\ell$~\cite{cadez}.  
We assume that this mass (computed from {\it apparent} horizons) 
is close to the total intrinsic mass of the black holes
(which is given by a knowledge of the spin---here zero---and the area of the 
{\it event} horizon).
The
binding energy, $\BE$, can be computed as the difference of the ADM mass 
observed at infinity and the sum of the horizon masses:
\be
\BE = M_{\ADM} - M_{\rm AH} - M'_{\rm AH}.
\ee
For Brill-Lindquist data $M_{\ADM} = m + m'$, so that
\be
\BE = -\frac{Gmm'}{\ell} +\rmO(\ell^{-2})\, ,
\label{eq:be_brill_lind}
\ee
which is the Newtonian result.

%----------------------------------------------------------------------
%
%
%
%----------------------------------------------------------------------
\subsection{Binding Energy in Superposed Kerr-Schild Data}

We now calculate the binding energy in superposed Kerr-Schild data (set
according to our conformal transverse-traceless approach) for a
non-moving Schwarzschild black hole at the origin, and a second such
hole at coordinate distance $\ell$ away. ($\ell$ is measured in the
flat space associated with the data construction.) We compute the area
of the hole at the origin to first order and find that the Newtonian
binding energy already appears in the the background data prior
to solving the constraints. Thus, we have an argument justifying the
result noted at the end of \sref{sec:boundary}: solving the elliptic
constraint equations leads to small corrections to the
Kerr-Schild background data.

Let both holes be placed on the $z$-axis; the first
hole with mass parameter $m$ at the origin, and a second hole with
mass parameter $m'$ at $z=\ell$.  The holes are well separated, and
we expand all quantities about the origin in powers 
of $\epsilon \equiv m'/\ell$ with
$\epsilon \ll 1$.  Using Schwarzschild coordinates labeled 
$(r, \theta, \phi)$ (cf.\ \eref{eq:1}--\eref{eq:6} for $a=0$), 
the background metric tensor is
\beq
\tilde g_{rr} &=& 1 + \frac{2m}{r} + 2\epsilon\cos^2\theta,\\ 
\tilde g_{r\theta} &=& -2\epsilon r \sin\theta \cos\theta,\\
\tilde g_{\theta \theta} &=& r^2 + 2\epsilon r^2 \sin^2\theta,\\
\tilde g_{\phi \phi} &=& r^2\sin^2\theta,
\eeq
with all other components zero.  The extrinsic curvature of the second
hole, ${}_2 K_{ab}$, is of $\rmO(\epsilon^2)$ at the origin, and we have
simply $\tilde K_{ab}={}_1 K_{ab}$.
%\beq
%\tilde K_{rho\rho} &=& -c_1 \frac{\rho+M}{\rho^3},\\
%\tilde K_{\theta\theta} &=& c_1,\\
%\tilde K_{\rho\rho} &=& c_1 \sin^2\theta,
%\eeq
%where
%\be
%c_1 \equiv 2M\sqrt{\frac{\rho}{\rho+2M}}.
%\ee
Similarly, the trace of the extrinsic curvature is $\tilde K = {}_1 K$.
%\be
%\tilde K = {}_1 K = c_1 \frac{(\rho+3M)}{\rho^2(\rho+2M)}.
%\ee
Finally, the non-zero components of $\tilde A_{ab}$ are
\beq
\tilde A_{r r} &=& -\frac{2c_2}{r^2}
           \lp 1 + 2\frac{m}{r} + 2\epsilon \cos^2\theta \rp,\\
\tilde A_{r \theta } &=&  \epsilon \frac{c_2}{r}\sin\theta \cos\theta,\\
\tilde A_{\theta \theta } &=&  c_2 \lp 1 + 2\epsilon \sin^2\theta \rp,\\
\tilde A_{\phi \phi} &=&  c_2\sin^2\theta.
\eeq
where
\be
c_2 \equiv  \frac{2M}{3} \sqrt{\frac{\rho}{\rho+2M}} \frac{(2r+3m)}{(r+2m)}.
\ee
While $\tilde K_{ab}$ is not a function of $\epsilon$, and hence
contains no information about the second hole, perturbative quantities
do appear in $\tilde A_{ab}$.  This perturbation in $\tilde A_{ab}$ arises
because we sum the mixed-index components of ${}_AA^c_b$, and because the
full background metric,
involving terms from each hole,
is involved in the symmetrization in~\eref{eq:ks_super}.

To calculate the binding energy we first find the apparent horizon area
of the local hole.   For a single Schwarzschild hole, the horizon
is spherical and located at $\rho_{\HOR} = 2m$;  the area of the horizon
is $16\pi m^{2}$.  The effect of the second hole is to distort the
horizon along the $z$-axis connecting them, and we define a trial apparent 
horizon surface as $f=0$, where
\be
   f = \rho - 2m - \sum_{l} a_{l} P_{l} (\cos \theta)\,.
\label{eq:f}
\ee
The expansion of $f$ in Legendre polynomials, $P_{l}$,
expresses the distortion of the local horizon away from the zero-order
spherical result.
This expansion includes a term describing a constant
``radial" 
offset in the position of the apparent horizon, $a_{0}P_{0}$.
This and the other terms defining the surface have the expected
magnitude, %$a_l=  \rmO \left(\frac{mm'}{\ell} \right)$.
$a_l=  \rmO (m\epsilon)$.
We solve for the horizon by placing this expression for the
surface into the apparent horizon equation
\begin{equation}
   \nabla_{i}s^{i} + A_{ab} s^{a}s^{b} - \frac{2}{3} K = 0,
\label{eq:app_horizon}
\end{equation}
where $s^{i}$ is the unit normal to the trial surface
\be
s_{i} = \frac{f_{,i}}{\sqrt{g^{ab}f_{,a}f_{,b}}}.
\ee

%\be
%   F_0(\rho) = \frac{2}{\rho} \frac{1-\frac{2m}{\rho}}{1+\frac{2m}{\rho}},.
%\label{30}
%\ee

The apparent horizon equation is solved to first order, $\rmO(\epsilon)$.
One must evaluate the equation at the new
(perturbed) horizon location.  Let $F$ represent the left-hand side of 
the apparent horizon equation (\eref{eq:app_horizon}), $\rho_0=2m$ is
the horizon surface of the single, unperturbed hole, and 
$\rho_\HOR(\theta)$ is the new perturbed horizon.  We expand $F$
to first order as
\be
F(\rho_\HOR(\theta)) = F_0(\rho_0) 
             + \frac{\partial F}{\partial \rho}\biggr|_{\rho_0} \sum a_{l}
   P_{l}\, = 0.
\label{31}
\ee
Solving (\eref{eq:app_horizon}), the 
only nonzero coefficients in Eq.(\ref{eq:f})
are $a_0=mm'/(3\ell),$ $a_2= -mm'/(2\ell)$.
Integrating the determinant of the perturbed metric over the horizon 
surface, $\rho = 2m + \sum_{l} a_{l} P_{l} (\cos \theta)$, we find
the area of the apparent horizon to be
\be
   A_\HOR = 16\pi \lp m + \frac{mm'}{2\ell}\rp^{2} 
            + \rmO\lp m^2(m'/\ell)^2\rp,
\label{32}
\ee
corresponding to a horizon mass of $M_\HOR = m + mm'/(2\ell)$ to Newtonian
order, ie to order $\rmO(\epsilon) = \rmO(\ell^{-1})$. 

In this nonmoving case the total ADM mass is just $M_\ADM = m + m'$. 
This leads to the Newtonian binding energy at this order
\be
\BE = -\frac{mm'}{\ell}.
\label{be}
\ee

Because we work only to lowest order in $\epsilon$, \eref{be} had to result
in an expression of $\rmO(m\epsilon)$, but it did not have to have
a coefficient of unity.
Both the conformally flat and conformally Kerr-Schild data
contain the Newtonian binding energy.
However, this result is obtained in the superposed {\it background} 
Kerr-Schild 
metric, while the Brill-Lindquist and \v Cade\v z data give
the correct binding energy only after solving the elliptic constraints.
This is consistent with the small corrections introduced by $\phi$ and
$w^i$ ($\phi \sim 1$, $|w^i| \ll 1$) in the solved Kerr-Schild data
(see \sref{sec:boundary}).
This fact---that for a superposed Kerr-Schild background the solution of the 
full elliptic problem modifies the data (and the mass/angular momentum 
computations) only 
slightly---demonstrates how powerful this choice of data can be.

Furthermore, the Newtonian form of the binding energy ($\epsilon \ll
1$) means the correct classical total energy is found for orbiting
situations.  If the holes have nonrelativistic motion, their individual
masses are changed by order $\gamma \approx 1+ \rmO(v^2) = 1 + \rmO(\epsilon)$.
The binding
energy, which is already $\rmO(\epsilon )$ and is proportional to the product
of the masses, is changed only at order $\rmO(\epsilon^2)$.  The ADM
mass, on the other hand, measures $\gamma m$, and $M_\ADM$ will be
increased by $m v^2/2$ (an  $\rmO(\epsilon)$ increase) for each hole,
leading to the correct Newtonian energetics for the orbit.

The apparent horizon is the only structure available to measure the
intrinsic mass of a black hole.   Complicating this issue is the intrinsic
spin of the black hole; the relation is between horizon area and 
{\it irreducible} mass:

\be
A_{\rm H} = 16 \pi m_{irr}^2 = 8 \pi m \lp m + \sqrt{(m^2 -a^2)}\rp
\label{eq:mirr}
\ee
As \eref{eq:mirr} shows, the irreducible mass is a function of both the 
mass and the spin, and in general we cannot specify the spin of the 
black holes. For axisymmetric cases Ashtekar's isolated horizon 
paradigm~\cite{Dreyer}
gives a way to measure the spin locally. We do not pursue the point here 
since we investigate generic and typically non-axisymmetric situations.

%----------------------------------------------------------------------
%
%
%
%----------------------------------------------------------------------
\subsection{Spin effects in Approximating Inspiral With Initial Data Sequences}
\label{sec:spineffects}

We have seen that the initial data contain the binding energy in a multiple
black hole spacetime.  This information can be used to deduce some
characteristics of the orbital dynamics, particularly the radius of 
the circular orbit, $\ell$, and the orbital frequency, $\omega$.
Given a sequence of initial data slices with
decreasing separations, we determine $\BE$ for each slice.
The circular orbit is found where
\be 
\frac{\partial \BE}{\partial \ell}\biggr|_{J} = 0.
\ee
The separation at the ISCO orbit, $\ell_\ISCO$, lies at the boundary between
binding energy curves which have a minimum, and those that do not. The
curve for the ISCO has an inflection point:
\be
\frac{\partial^2 \BE}{\partial \ell^2}\biggr|_{J} = 0.
\ee
The 
angular frequency is given by
\be
\omega_\ISCO = \frac{\partial \BE}{\partial J} .
\label{omega}
\ee

The attempt to model dynamical inspiral seems secure for large
separation ($\ell>15 m$), though
surprises appear even when the holes are very well separated.
For instance, \eref{32} above shows that
compared to the bare parameter values, the
mass increase is equal for the two holes in a
dataset. Thus the smaller hole is proportionately more strongly affected
than the larger one is. 
 
The physically measurable quantity in question is the frequency (at
infinity) \eref{omega} associated with the last orbit prior to the plunge, the
ISCO. This may be impossible to determine by the initial data set method.

To begin with, isolated black holes form a 2-parameter set (depending
on the mass parameter $m$, and the angular momentum parameter,
$j=ma$).  For isolated black holes without charge 
the parameters $\{m, j\} $ uniquely
specify the hole. They are equal, respectively, to the
physical mass and angular momentum.  Every method of constructing
multiple black hole data assigns parameter values ${}_Bm$ and ${}_Bj$ to
each constituent ${}_B(hole)$ in the data set.  

There is substantial ambiguity involving spin and mass in setting the
black hole data. One must consider the evolutionary
development of the black hole area and spin. This is a real physical 
phenomenon which contradicts at some
level the usual assumption of invariant mass and spin. A related 
concern arises because it is only the {\it total} ADM angular momentum that
is accessible in the data, whereas one connects to particle motion via
the {\it orbital} angular momentum.

Consider the behavior of the individual black hole spin and mass 
in an inspiral. For widely separated holes, because
the spin effects fall off faster with distance than the dominant
mass effects do, we expect the spin to be approximately conserved in an 
inspiral. Therefore it should also be constant across the initial data sets
representing a given sequence of orbits. But when the holes approach closely,
the correct choice of spin parameter becomes problematic also.  

Newtonian arguments demonstrate some of the possible spin effects. In every
case they are {\it a priori} small until the orbits approach very
closely. However, at estimates for the ISCO, the effects begin to be
large, and result in ambiguities in setting the data (see Price and 
Whelan~\cite{P+W}).
We will consider these effects in decreasing order of their magnitude.

For two holes, each of mass $m$ in Newtonian orbit with a total separation of
$\ell$, the orbital frequency is

\be 
m \omega = \sqrt 2 (m/\ell)^{(3/2)}.  
\label{35c} 
\ee 

From recent work by Pfeiffer et al.~\cite{Pfeiffer}, the estimated 
ISCO frequency is of order $m\Omega=0.085$, corresponding to $\ell \approx
6.5\, m$ in this Newtonian approach.

To compare this frequency, \eref{35c}, 
to an intrinsic frequency in the problem, we take the lowest
(quadrupole) quasi-normal mode of the final merged black hole (of mass
$2m$) which has frequency $2m\omega_0 \approx 0.37$; the quadrupole
distortion is excited at twice the orbital frequency. (We are using the
values for a Schwarzschild black hole in this qualitative analysis.) The
driving frequency equals the quasi-normal mode frequency when
$\ell \approx 4\, m$, as might be expected.  

To consider effects linked to the orbital motion on the initial
configurations, we can first treat the effect of imposing corotation.
While we show below that corotation is not physically enforced except
for very close orbits, it is a fact that certain formulations, for
instance versions of the ``thin sandwich" with a helical Killing vector,
require corotation in their treatment. For any particular
initial orbit, corotation is certainly a possible situation. 

In corotation, then, with \eref{35c}, for each hole:
\be 
J = m a = I\omega = 4 m^2 (m \omega).  
\label{35dd} 
\ee 
The result for the moment of inertia $I=4m^3$ is the Schwarzschild 
value~\cite{MTW,membrane}. Thus
\be 
a = 4m \sqrt{2}  (m/\ell)^{(3/2)}.  
\label{35e} 
\ee 
Assume $a/m \ll 1$, and compute the area of this black hole~\cite{MTW}:
\be 
A= 8 \pi m(m+\sqrt{m^2-a^2})\\ 
\approx 16\pi m^2 (1- (a/m)^2/4).
\label{35ff} 
\ee 
The horizon mass computed from this area is 
\be 
\sqrt{A/(16\pi)} \approx m(1-4(m/\ell)^3).
\label{35f} 
\ee 

At our estimate of the ISCO orbit, $\ell_\ISCO\approx 6\, m$, this effect is
of order of 10\% of the Newtonian binding energy, distinctly enough to affect 
the location of the ISCO.(At $\ell_\ISCO\approx 6\, m$, $a/m \approx 0.3$
for corotation).

Two more physical effects are not typically considered in setting data.
They are {\it frame dragging}, and {\it tidal torquing}. Within our
Newtonian approximations, we will find that these effects are small, but
not zero as the orbits approach the ISCO. 
In full nonlinear gravity these effects could be substantial precisely
at the estimated ISCO. 

The frame dragging is the largest dynamical
effect. The orbiting binary possesses a net angular momentum. For a
rotating mass (here the complete binary system) the frame dragging
angular rate is estimated as the rotation rate times the
gravitational potential at the measurement point~\cite{MTW}. Hence 
\be 
m\Omega_{\rm{drag}}= m\omega\lp \frac{2m}{\ell}\rp 
         \approx \lp\frac{m}{\ell}\rp^{\frac{5}{2}}
       \approx \frac{a}{4m}.
\label{35gg} 
\ee 
This is  $a/m$ of order 1\% at $\ell=10m$;  of order 4\% at $\ell=6m$.

The tidal torquing and dissipative heating of the black holes can be
similarly estimated. 
As the two holes spiral together, the
tidal distortion from each hole on the other will have a frequency
which is below, but approaching the quasi-normal frequency. Just as for
tidal effects in the solar system, there will be lag in the phase angle
of the distortion, which we can determine because the lowest
quasi-normal mode is a dissipative oscillator, driven through the tidal 
effects at twice the orbital frequency:  
\be 
\ddot q +2 \gamma \dot q + \omega_0^2 q= F(\omega).  
\label{35d} 
\ee 
Here $m^2q$ is the quadrupole moment of the distorted black hole. The
parameter $\gamma$ is (for a Schwarzschild hole of mass $2m$)) about 
$2m \gamma = 0.089$. In \eref{35d} the driving acceleration $F(\omega)$
is identified with the tidal distortion acceleration. We evaluate it at zero
frequency:
\beq 
q &=& F(\omega=0)/\omega_0^2\\
&\approx & \frac{m}{\ell^3}.  
\label{36} 
\eeq 
The lagging phase, for driving frequency $2\omega \ll \omega_0$, is easily computed
to be
\beq 
\phi &\approx& 4 \gamma \frac{\omega}{\omega_0^2}\\
&=& 4\lp\frac{\gamma}{\omega_0}\rp \lp\frac{\omega}{\omega_0}\rp
\label{37} 
\eeq 
This lagging tidal distortion will produce a tidal torque on the black
hole, which we can approximate using a combination of Newtonian and
black hole ideas. The most substantial approximation is that the torque
arises from a redistribution of the mass in the ``target" black hole, of
amount $\Delta m  =  m m^2q = m (m/\ell)^3$. This mass has separation
$\approx 4m$. Thus the torque on the hole is
\begin{eqnarray} 
\tau &=& \sin \phi \times (\mbox{lever arm}) \times \Delta F \\ \nonumber
&=& \sin \phi \times (4m) \times (\Delta m 2m^2 / \ell^3) \\ \nonumber
&=&  8 \sin \phi m(m/\ell)^6\\ \nonumber
&\approx & 32 (\gamma/\omega_0)(\omega/\omega_0)m(m/\ell)^6\\ \nonumber
&\approx& 60 m(m/\ell)^{15/2}. 
\label{38} 
\end{eqnarray}  

What is most important is the effect of this torque on the angular
momentum of the hole over the period of time it takes the orbit to
shrink from a very large radius. To accomplish this, we use the inspiral
rate (calculated under the assumption of weak gravitational radiation
from the orbit; see~\cite{MTW}): 
\be 
\frac{d\ell}{dt} = -\frac{128}{5} \lp \frac{m}{\ell}\rp^3
\label{39} 
\ee 
Thus 
\begin{eqnarray} 
\frac{dJ}{d\ell} &=& \tau  \frac{dt}{d\ell} \\
&=& -\frac{5\tau}{128} \lp \frac{m}{\ell}\rp^{-3} \\
&\approx& -2  m \lp \frac{m}{\ell}\rp^{9/2}, 
\label{40}
\end{eqnarray}
and 
\be 
J(\ell) \approx m^2 \left(\frac{m}{\ell}\right)^{7/2};
\label{40x} 
\ee 
assuming that there is minimal mass increase from the associated
heating (which we discuss just below), this identifies the induced spin
parameter $a= m(m/\ell)^{7/2}$ for an inspiral from infinity.

The estimate $a= m(m/\ell)^{7/2}$ for an inspiral from infinity assumes the
mass of the hole has not changed significantly in the inspiral.  By
considering the detailed behavior of the shear induced in the horizon
by the tidal perturbation, the growth in the black hole mass can be
estimated~\cite{membrane} as 
\begin{eqnarray} 
\frac{dm}{dt} &=& \omega \frac{dJ}{dt},
\label{41}
\end{eqnarray}
leading to a behavior 
\be 
\Delta m(\ell) \approx 5 m \lp \frac{m}{\ell}\rp^{5};
\label{42} 
\ee 
Consequently, the change in mass can be ignored until the holes are quite
close. However, the point is that these Newtonian estimates lead to
possible strong effects just where they become unreliable, and just
where they would affect the ISCO. 

These results are consistent with similar ones of Price and Whelan~\cite{P+W},
who estimated tidal torquing using a derivation due to
Teukolsky~\cite{TeukThesis}. That derivation assumes the quadrupole
moment in the holes arises from their Kerr character, which predicts
specific values for the quadrupole moment, 
as a function of angular momentum parameter $a$. 

Finally we consider an effect on binding energy shown by Wald and also by Dain.
Wald directly computes the force for stationary sources with 
arbitrarily oriented spins. He considered a small black hole as a 
perturbation in the field of a large hole. The result found~\cite{WaldPRD} was

\be
\BE = - \frac{mm'}{\ell} - \lp \frac{ \vec{S} \cdot \vec{S'} - 
        3(\vec{S} \cdot \hat{n})(\vec{S'} \cdot \hat{n})}{\ell^3} \rp.
\label{BEWald}
\ee

\noindent Here, $\vec{S}$, $\vec{S}'$ are the spin vectors of the
sources and $\hat{n}$ is the unit vector connecting the two sources.
Dain~\cite{Dain}, using a definition of intrinsic mass that differs
from ours, finds binding energy which agrees with Wald's \eref{BEWald}
at $\rmO(\ell^{-3})$. This is discussed further in Section
\ref{sec:physresults}.

%----------------------------------------------------------------------
%
%
%
%----------------------------------------------------------------------
\section{Numerical Results}

We now turn to computational solutions of the constraint equations
to generate physical data using the superposed Kerr-Schild data.
We first discuss the computational code and tests, as well as
some of the limitations of the code.  Finally, we consider
physical conclusions that can be drawn from the results.

%----------------------------------------------------------------------
%
%
%
%----------------------------------------------------------------------
\subsection{Code Performance}

The constraint equations are solved (\eref{ell_eqs}) with an accelerated 
SOR solver~\cite{NumRecepies}. The solution is iterated until the $L_{2}$ 
norms of the residuals 
of the fields are less than $10^{-10}$, far below truncation error.
Discrete derivatives are approximated with second order, centered derivatives.
We are limited to fairly small domains, e.g., $x^i \in [-12 m, 12m]$ 
for a typical $m/8$ resolution using $193^3$ points.

To verify the solution of the discrete equations, we have examined the 
code's convergence in some detail.   The constraints have known analytical
solutions---they should be zero---which allows us to determine the code's
convergence using a solution calculated at two different resolutions.
Let $S_1$ be a solution calculated with resolution $h_1$, and
$S_2$ be a solution calculated with $h_2$, then the convergence
factor $c_{12}$ is
\begin{eqnarray}
c_{12} = \frac{\log\left(\frac{||S_{1}||}
{||S_{2}||}\right)}{\log\left(\frac{h_{1}}{h_{2}}\right)}
\label{eq:convCalc}
\end{eqnarray}

\begin{figure}
\begin{center}
\includegraphics[width=3.in,angle=270]{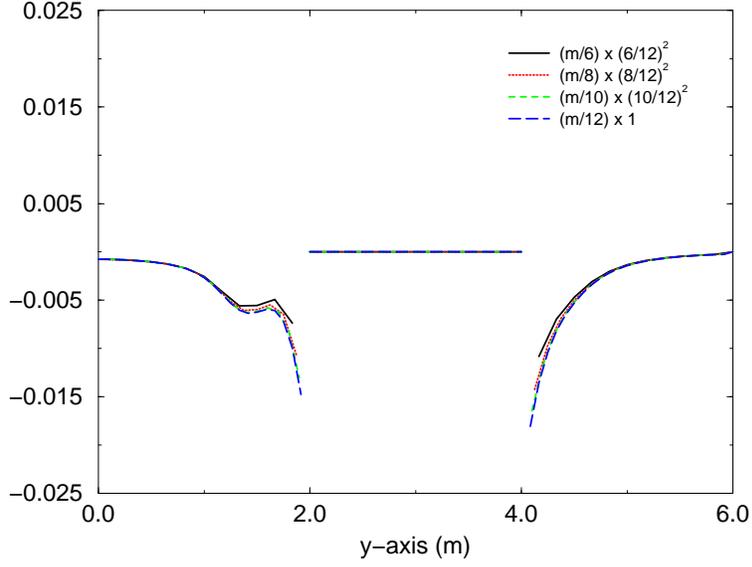}
\end{center}
\caption{The Hamiltonian constraint (units $m^{-2}$) along $y$-axis after
solving the elliptic equations for 4 different levels of resolution. The
constraints are rescaled by the ratio of the resolutions squared,
showing second
order convergence. The two non-spinning, instantaneously stationary
holes of $m=1$ are positioned at $\pm3$ on the $y$-axis.}
\label{fig:ham2}
\end{figure}

\begin{figure}
\begin{center}
\includegraphics[width=3.in, angle=270]{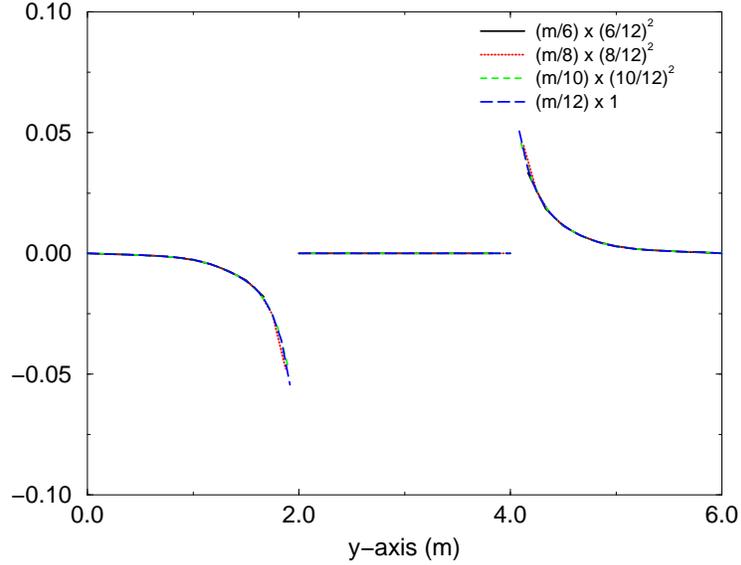}
\end{center}
\caption{$y$-component of momentum constraint (units $m^{-2}$) along the $y$-axis after
solving the elliptic equations for 4 different levels of resolution,
showing second order convergence. The background physical
situation is the same as in
Figure (\ref{fig:ham2}). The other momentum constraint components evaluated
on this axis are zero by symmetry, both analytically, and computationally
($\rmO(10^{-12})$).}
\label{fig:momy2}
\end{figure}

\begin{figure}
\begin{center}
\includegraphics[width=3.in, angle=270]{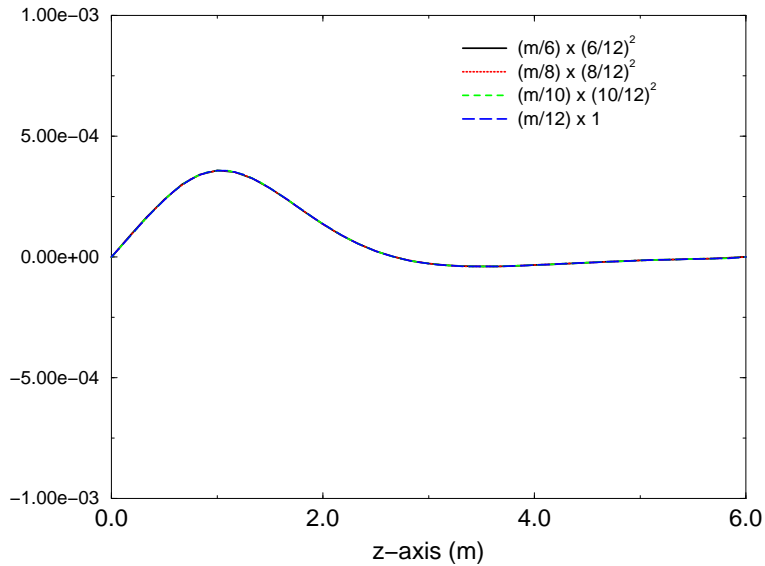}
\end{center}
\caption{$z$-component of momentum constraint (units $m^{-2}$) along the $z$-axis after
solving the elliptic equations for 4 different levels of resolution,
showing second order convergence.
Other components of the momentum constraint evaluated along this line
are zero by symmetry, both
analytically and computationally ($\rmO(10^{-12})$).
The background physical situation
is the same as in
Figure (\ref{fig:ham2}). The behavior of the $x$-momentum constraint
along the $x$-axis is identical to this Figure, as required by the
symmetry of the problem.}
\label{fig:momz2}
\end{figure}

We constructed a conformal background spacetime with two $m=1$ non-spinning
black holes separated by $6m$ on the $y$-axis.
The elliptic equations were then solved on grids with resolutions of
$m/6$, $m/8$, $m/10$ and $m/12$. 
Tables \ref{tab:TableHam}--\ref{tab:TableMomX} show the convergence factors
as a function of resolution for the Hamiltonian constraint and 
the $x$-component of the momentum constraint, ${\cal C}^x$.   The 
convergence for ${\cal C}^y$ is nearly identical to ${\cal C}^x$, and
as the $y$-axis is an axis of symmetry, ${\cal C}^z$ is identical to 
${\cal C}^x$.
Figures \ref{fig:ham2}--\ref{fig:momz2} show the convergence behavior of the
constraints along coordinate lines. The constraints
calculated at lower resolutions are rescaled to the highest resolution by
the ratio of resolutions squared. We see second order for all components
with the exception of the points nearest to the inner boundary.

\begin{table}
\begin{tabular}{|l|ccc|}\hline
\multicolumn{4}{|c|}{Convergence $(c_{ab})$}\\
~~ & ~~$a=m/6$ & ~~$a=m/8$ & ~~$a=m/10$ \\ \hline
$b=m/8$  & 1.70 & ~~ & ~~ \\
$b=m/10$ & 1.77 & 1.86 & ~~ \\
$b=m/12$ & 1.79 & 1.86 & 1.85 \\ \hline
\end{tabular}
\vspace{.5in}
\label{tab:TableHam}
\caption{Convergence data for the Hamiltonian constraint,
${\cal C}^0$, for a solution with two $m=1$, non-spinning holes
at $x^i=(0,\pm 3m,0)$ in the conformal background, and outer boundaries 
at $x^i=\pm 6m$.  The solution was calculated at resolutions $m/6$, 
$m/8$, $m/10$, and $m/12$.  The $L_2$ norms of ${\cal C}^0$ were calculated
over the entire volume of the domain
using a mask of radius $1m$ around each hole, while the computational mask
has a radius of approximately $0.75~m$.  This larger mask was used to
compensate for the slight difference of physical location of the mask
at different resolutions.  The norms are as follows: 
$||{\cal C}^0(m/6)||_2 = 0.00389054$,
$||{\cal C}^0(m/8)||_2 = 0.00238321$,
$||{\cal C}^0(m/10)||_2 = 0.00157387$ and
$||{\cal C}^0(m/12)||_2 = 0.00112328$.
}
\end{table}

\begin{table}
\begin{tabular}{|l|ccc|}\hline
\multicolumn{4}{|c|}{Convergence $(c_{ab})$}\\
~~ & ~~$a=m/6$ & ~~$a=m/8$ & ~~$a=m/10$ \\ \hline
$b=m/8$ & 1.93 & ~~ & ~~ \\
$b=m/10$ & 1.99 & 2.06 & ~~ \\
$b=m/12$ & 1.99 & 2.03 & 1.99 \\ \hline
\end{tabular}
\vspace{.5in}
\label{tab:TableMomX}
\caption{Convergence data for the $x$-component of the 
momentum constraint, for the same configuration as Table~\ref{tab:TableHam}.
The norms of ${\cal C}^x$ are as follows:
$||{\cal C}^x(m/6)||_2 = 0.00541231$,
$||{\cal C}^x(m/8)||_2 = 0.00310937$,
$||{\cal C}^x(m/10)||_2 = 0.00196156$ and 
$||{\cal C}^x(m/12)||_2 = 0.00136514$.
Convergence factors were also calculated for ${\cal C}^y$ and ${\cal C}^z$,
and found to be essentially identical to the data shown here, and thus
are not given separately.
}
\end{table}

%\begin{table}
%\begin{tabular}{cc}\hline\hline
%\multicolumn{2}{c}{Momentum$_{y}$:}\\\hline\hline
%Resolution & L2 norm \\
%m/6 & 0.00551741 \\
%m/8 & 0.00315638 \\
%m/10 & 0.00198946 \\
%m/12 & 0.00138186
%\end{tabular}

%\begin{tabular}{lccc}\hline
%\multicolumn{4}{c}{Convergence $(c_{ab})$}\\
%~~ & a=6 & a=8 & a=10 \\ \hline
%b=8 & 1.94131925291174 & ~~ & ~~ \\
%b=10 & 1.99685617693102 & 2.06845572486942 & ~~ \\
%b=12 & 1.9973797316871 & 2.03715527437679 & 1.99884661903467 \\ \hline
%\end{tabular}
%\vspace{.5in}
%\label{tab:TableMomY}
%\end{table}

%\begin{table}
%\begin{tabular}{cc}\hline\hline
%\multicolumn{2}{c}{Momentum$_{z}$:}\\\hline\hline
%Resolution & L2 norm \\
%m/6 & 0.00541231 \\
%m/8 & 0.00310937 \\
%m/10 & 0.00196156 \\
%m/12 & 0.00136514
%\end{tabular}

%\begin{tabular}{lccc}\hline
%\multicolumn{4}{c}{Convergence $(c_{ab})$}\\
%~~ & a=6 & a=8 & a=10 \\ \hline
%b=8 & 1.9266263300294 & ~~ & ~~ \\
%b=10 & 1.98685396059259 & 2.0645008810491 & ~~ \\
%b=12 & 1.98719556241069 & 2.0301701166937 & 1.9881526569248 \\ \hline
%\end{tabular}
%\vspace{.5in}
%\label{tab:TableMomZ}
%\end{table}

%------------------------------------------------
\begin{table}
\begin{tabular}{|c|c|}
\hline \hline
Domain &  $M_{ADM}$ \\
\hline
$\pm$ 8 m & 1.942 m \\
$\pm$ 10 m & 1.964 m \\
$\pm$ 11 m & 1.974 m \\
$\pm$ 12 m & 1.980 m \\
\hline \hline
\end{tabular}
\caption{Total ADM Mass for two instantaneously stationary, non-spinning
holes separated by $6 m$ on a grid of discretization $\Delta x = m/8$ for
four different domain sizes.}
\label{tab:table1}
\end{table}
%------------------------------------------------

\begin{figure}
\begin{center}
\includegraphics[width=3.0in]{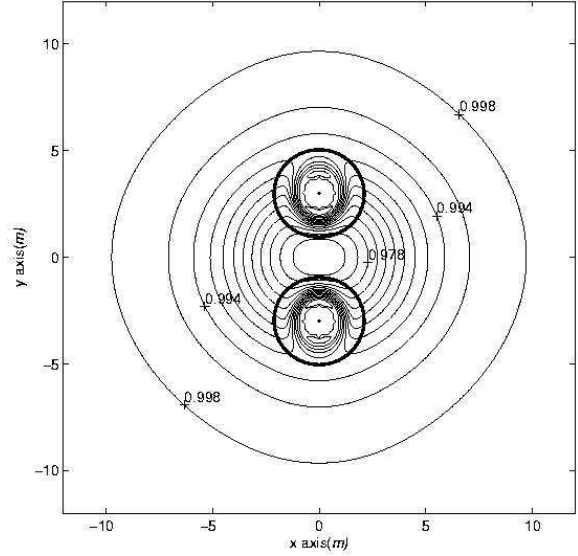}
\end{center}
\caption{
Contour plot of $\phi$ for two instantaneously stationary, non-spinning
holes of mass parameter $m = 1$.
The holes are separated by $6~m$ along the y-axis. The bold circles indicate
the apparent horizons.}
\label{fig:phi_zero}
\end{figure}

\begin{figure}
\begin{center}
\includegraphics[width=3.0in]{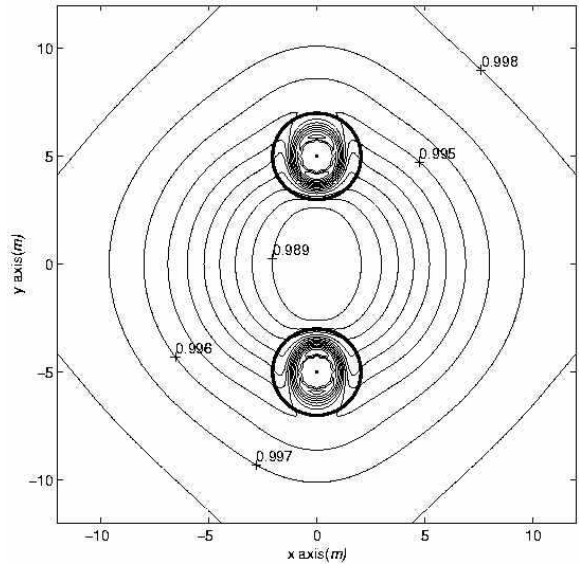}
\end{center}
\caption{
Contour plot representing the same configuration as \fref{fig:phi_zero}
 but with the holes separated by 
$10$~m along the y-axis. Compare to Figure 8 in ref.~\cite{Pfeiffer2}}.
\label{fig:phi_zero10}
\end{figure}

Solutions of elliptic equations are well-known to be dependent on all
boundary data.  The outer boundary is an artificial boundary, as the
the physical spacetime is unbounded.  Boundary data for this outer
boundary are derived from the asymptotic behavior of a single Kerr 
black hole.  On very large domains these conditions should closely
approximate the expected field behavior, but on small domains
these boundary data may only crudely approximate the real solution.
This error in the boundary data then contaminates the entire solution,
as expected for elliptic solutions.  Additional error arises in the
calculation of the ADM quantities, as spacetime near the outer boundary
does not approach asymptotic flatness.   
As an indication of the error associated
with the artificial outer boundaries, we calculated solutions with the
same physical parameters on grids of differing sizes.  The boundary
effects in the $M_\ADM$ are given in \tref{tab:table1}, and 
\fref{fig:phi_zero} shows a contour plot of $\phi$ for equal mass,
nonspinning, instantaneously stationary black holes with the outer
boundaries at $x^i = \pm 12m$.  As a further demonstration of
boundary effects in our solutions, \fref{fig:phi_zero10} shows
$\phi$ for a configuration examined by Pfeiffer {\it et al.}~\cite{Pfeiffer2}.
Their solution, shown in Fig.~8 of~\cite{Pfeiffer2}, was computed on a 
much larger domain via a spectral method~\cite{Kidder}.
Thus, while we achieve reasonable results,
it is important to remember that the boundary effects may be significant.
Moreover, we have only considered the effect of outer boundaries, while
errors arising from the approximate inner boundary condition have
not been examined.

\begin{figure}
\begin{center}
\includegraphics[width=3.in,angle=270]{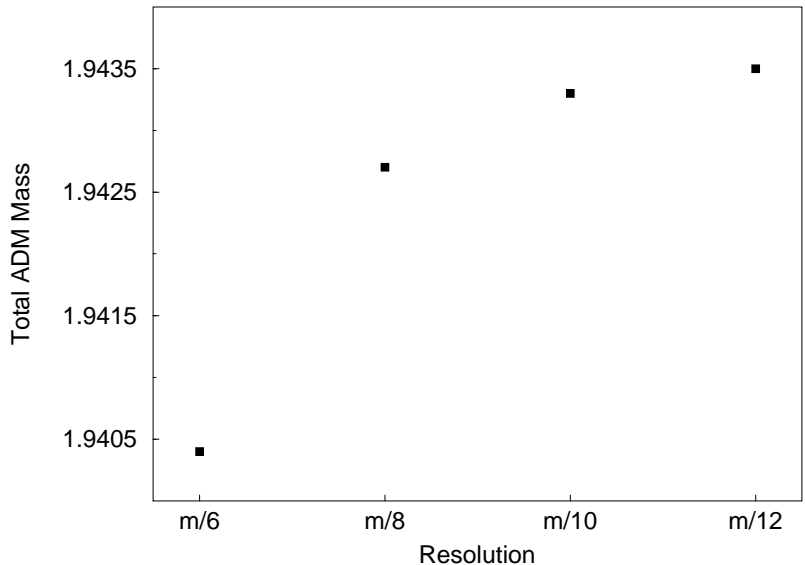}
\end{center}
\caption{The total ADM energy for two momentarily stationary non-spinning
black holes separated by $6 m$ at various resolutions. 
The results exhibit second order convergence.}
\label{fig:2bh_adm_mass}
\end{figure}

\begin{figure}
\begin{center}
\includegraphics[width=3.in,angle=270]{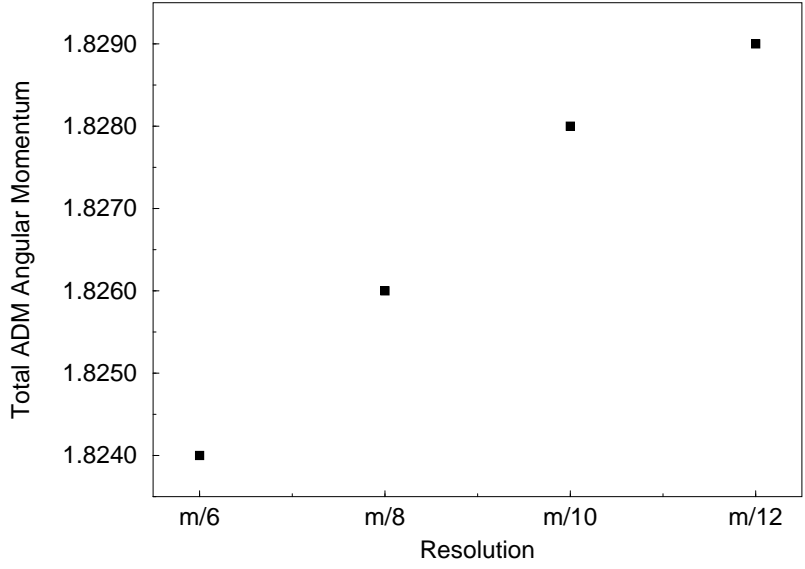}
\end{center}
\caption{The total ADM angular momentum for two non-spinning holes 
boosted in the $\pm$
x direction with $v=0.3162$ and separated by $6 m$ at various 
resolutions (background angular momentum $\tilde J^{ADM}_{12} = 2.0 m^2$).}
\label{fig:2bh_adm_angMom}
\end{figure}

\begin{figure}
\begin{center}
\includegraphics[width=3.in]{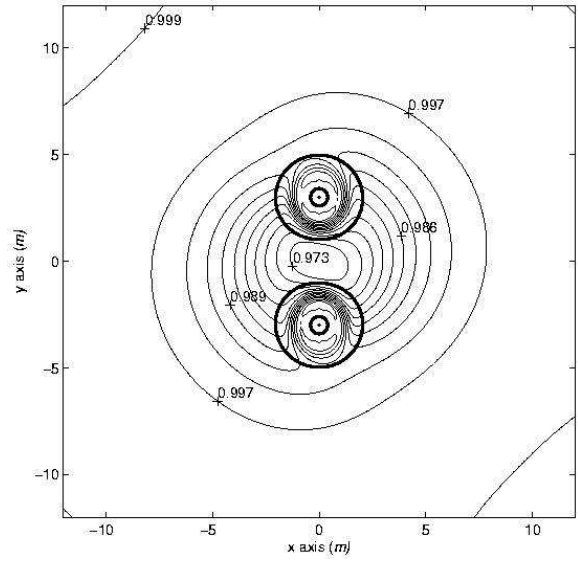}
\end{center}
\caption{Conformal factor $\phi$ for two instantaneously stationary holes 
separated by $6m$ 
with spin parameter $a=0.5$. The spins are parallel and pointed out of
the page. Compare to \fref{fig:phi_zero}. Also notice the boundary 
effect on the outermost contour, labeled $0.999$. }
\label{fig:phi_spinaligned}
\end{figure}

\begin{figure}
\begin{center}
\includegraphics[width=3.in]{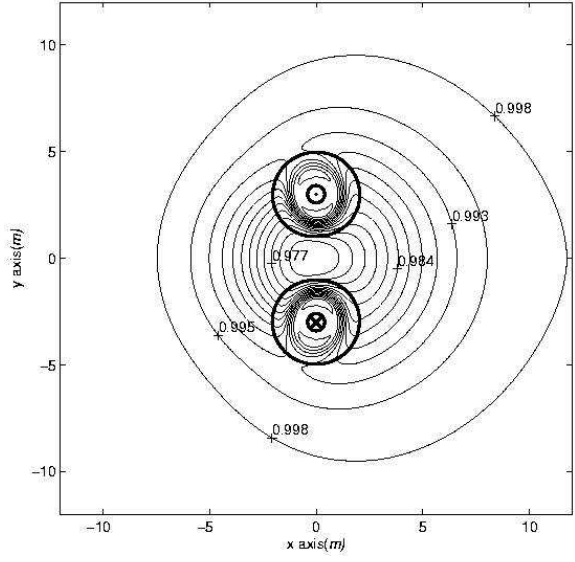}
\end{center}
\caption{Conformal factor $\phi$ for the same configuration 
as \Fref{fig:phi_spinaligned} except the spins are anti-parallel: 
the the spin of the hole at $(0,-3,0)$ points into the page.}
\label{fig:phi_spinantialigned}
\end{figure}

\begin{figure}
\begin{center}
\includegraphics[width=2.5in]{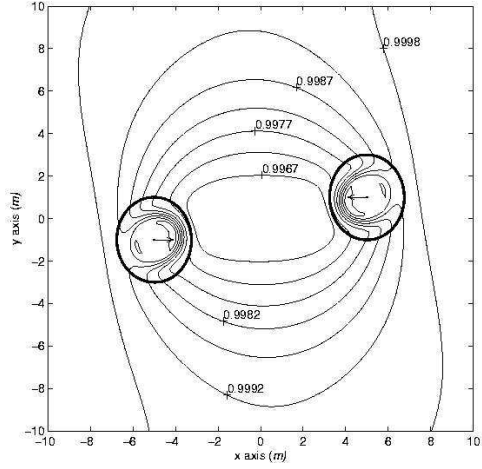}
\end{center}
\caption{$\phi$ for a grazing collision between
two equal mass, non-spinning holes. The holes are centered at $y = \pm 1 m$
and boosted toward each
other with $v_x = \mp 0.5c$, respectively. }
\label{fig:grazing}
\end{figure}

\begin{figure}
\begin{center}
\includegraphics[width=2.5in]{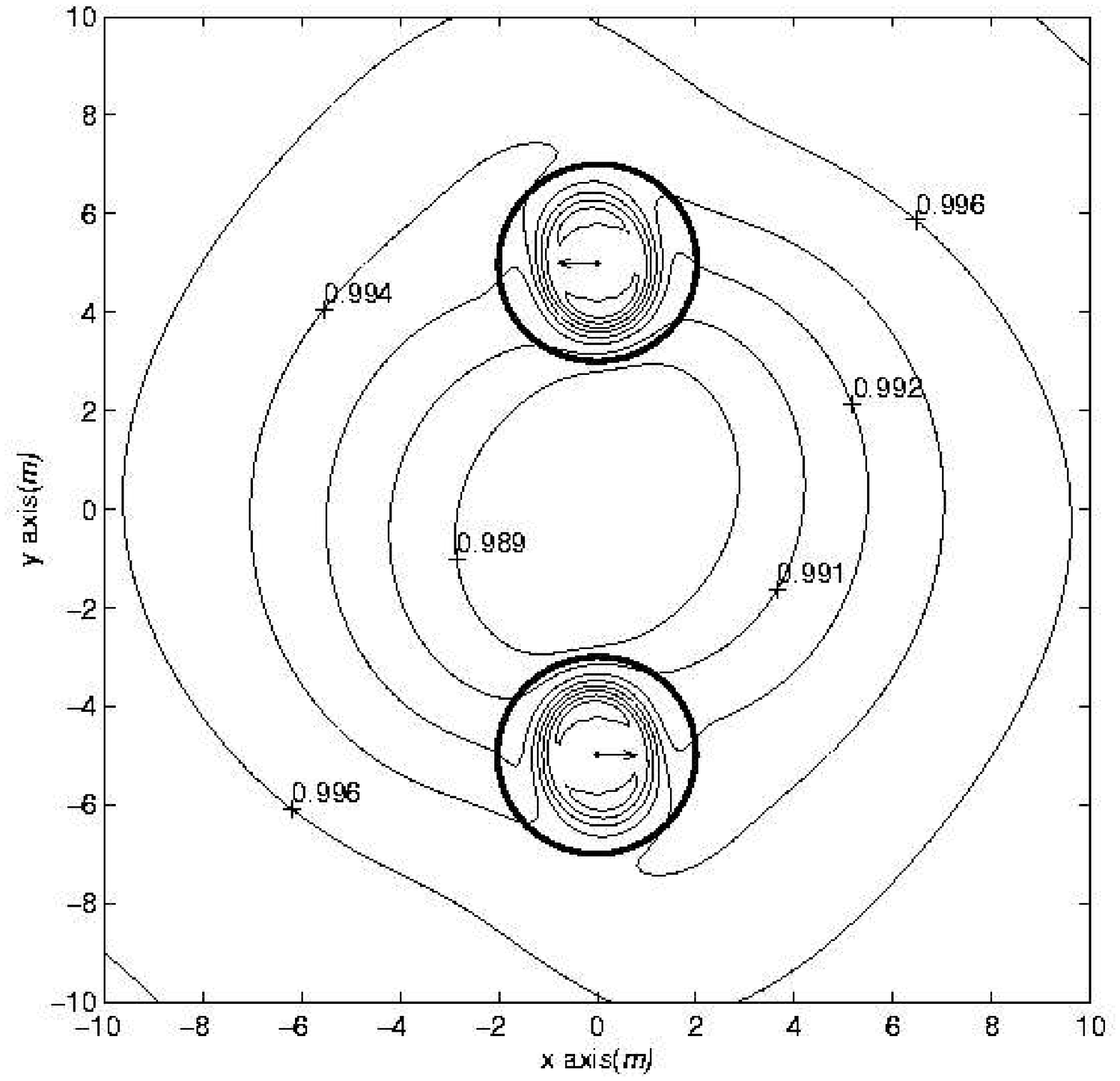}
\end{center}
\caption{$\phi$ for two non-spinning holes boosted 
perpendicularly to their separation. The holes are separated by $10 m$
and boosted 
with $v_x = \pm 0.196$, giving the system a background 
angular momentum of $\tilde J^{ADM}_{12} = 2.0 m^2$. 
The calculated $J^{ADM}_{12} = 1.91 m^2$ and 
$M_{ADM} = 1.970 m^2$. The Newtonian data correspond to an elliptic orbit
at apastron. }
\label{fig:circular}
\end{figure}

Other derived quantities also show convergence: \fref{fig:2bh_adm_mass} 
shows the ADM mass $M_{\ADM}$ for two nonspinning black holes at $6m$ 
separation, and different resolutions. 
The fit is 
\be 
M_{ADM} = \lp 1.941 + 0.067 \lp \frac{\Delta x}{m}\rp  - 0.422 
\lp \frac{\Delta x}{m} \rp^2 \rp m.
\label{eq:42S} 
\ee  
showing good second order convergence. The angular momentum calculation
is less robust, but exhibits approximately first order convergence. 
The fit to Figure \ref{fig:2bh_adm_angMom}, which shows $J^{ADM}$ for
two nonspinning holes with orbital angular momentum, gives:

\be 
J^{ADM}_{12} = \lp 1.837 - 
0.121 \lp \frac{\Delta x}{m}\rp + 
0.237 \lp \frac{\Delta x}{m} \rp^2 \rp \epsilon_{12}m^2.
\label{eq:42T} 
\ee 
Compare this to the angular momentum computed 
for the background: $\tilde J^{ADM}_{12} = 2.0 m^2$.

%----------------------------------------------------------------------
%
%
%
%----------------------------------------------------------------------
\subsection{Physics Results}
\label{sec:physresults}

The small computational domain does not negate the 
utility of these solutions as initial data for the time-dependent
Einstein equations. For instance, Figures \ref{fig:grazing} and 
\ref{fig:circular} show data for grazing and elliptical orbits.
They are currently being incorporated into the Texas binary black hole
evolution code. While the small domains do mean that 
that our data do not represent the best asymptotically flat results
available from this method, we can still verify some of the qualitative
analytical predictions of the previous section. 
In particular, Figures  \ref{fig:phi_zero}, \ref{fig:phi_spinaligned}, 
 and \ref{fig:phi_spinantialigned} 
show the conformal factor $\phi$
for holes instantaneously at rest at a separation of $6m$. 
In Figure \ref{fig:phi_zero} they are
non-spinning; in  Figures
 \ref{fig:phi_spinaligned} and
\ref{fig:phi_spinantialigned}, each has Kerr parameter $a=0.5m$. 
In one case (\fref{fig:phi_spinaligned}) the spins are aligned; 
in the other (\fref{fig:phi_spinantialigned}) they are antialigned. 
\tref{tab:table2} gives
the values of the apparent horizon area of each hole, the ADM mass, and the
binding energy fraction for these configurations. 
The binding energy is consistent with the
analytic estimates of  Wald~\cite{WaldPRD} in Section \ref{sec:spineffects}.

Wald's computation of the binding energy for spinning holes,
\eref{BEWald}, gives for parallel or antiparallel spins orthogonal to
the separation (as in our computational models):

\be
\BE = -  \lp \frac{mm'}{\ell} + \frac{ \vec{S} \cdot \vec{S'}}{\ell^3} \rp
\label{BEwald}
\ee
with oppositely directed spins showing {\it less} binding energy.
For our $S = 0.5~m^2$, $\ell = 6~m$ configuration, this is a 
change of order $\rmO(2 \times 10^{-3})$ between the parallel and the
antiparallel cases. For the spinning cases we compute a 
change in binding energy between parallel and antiparallel 
spins of roughly half that,
with the correct sign. This rough 
correspondence to the analytic result is suggestive. However, the nonspinning 
case deviates from the expectation that its binding energy 
should be between that 
of the spinning cases. Based on the scatter in the binding energies 
shown here, we estimate that we have achieved about $3 \%$ accuracy in the 
binding energy. With the accuracy of our solution and the size of our domain,
we are unable to present a clearer dependence of binding energy on spin.

\begin{table}
\begin{tabular}{l || l | l | l | l}
\hline \hline
 & $A^{\dag}_{AH}$ & $M_{ADM}$ & $M^{\dag}_{AH}$ & Binding Energy \\
\hline \hline
Parallel Spin & 53.20 & 1.973 & 1.065 & -0.157 = -0.147 $\times M_{AH}$ \\
Antiparallel Spin & 53.17  & 1.974 & 1.065 & -0.156 = -0.146 $\times M_{AH}$ \\
Zero Spin & 57.10 & 1.980 & 1.066 & -0.151 = -0.142 $\times M_{AH}$ \\
\hline \hline
\end{tabular}
\caption{$M_{ADM}$, $A_{AH}$ and associated quantities 
calculated for two holes with $m = 1.0$ on a grid $(24~m)^3$ with 
resolution $\Delta x = m/8$. \\
${}^\dag$: Quantity for a single hole.\\}
\label{tab:table2}
\end{table}

%----------------------------------------------------------------------
%
%
%
%----------------------------------------------------------------------
\section{Outlook}

To an extent, the difficulty in setting data will become less relevant,
as good evolutions are eventually achieved. Then data can be set for
initial configurations with very large separation, and the subsequent
evolution will tell us the future dynamics. In the shorter term, the
iBBH program of Thorne and collaborators~\cite{Brady} 
will give us an indication of
the evolution of the black hole parameters in the inspiral, and will
allow a closer identification of the corresponding initial data
sequence. To point out a couple of additional physical effects, 
note that, besides the historical component associated
with a lagging tidal distortion, there is the familiar fact that most
data setting methods are incapable of accounting for the previously
emitted gravitational radiation. One can then expect that data
describing hyperbolic encounters will be more accurate than data sets
describing circular motion. This is because, in hyperbolic encounters, which are
set as distant initial configurations, the radiation is more planar,
and confined to near each hole.
 The radiation should then both be better defined, and should 
have less effect on the subsequent evolution than in the more distorted 
orbiting data set.  In any case, the
understanding of these problems is extremely significant in
understanding the physical content of the configurations we must solve
to provide waveforms for the new generation of gravitational wave
detectors.
For more accurate computational results, we are undertaking a both
multigrid approach~\cite{Klasky}, and a spectral approach~\cite{Pfeiffer2, GGB1}
and expect to have extended results soon comparable to those
of~\cite{Pfeiffer2,Kidder}. An ultimate goal of such a solver is to be 
able to carry 
out elliptic solutions at every integration time step, to enable fully 
constrained evolutions.

%----------------------------------------------------------------------
%
%
%
%----------------------------------------------------------------------
%\subsection{Apparent Horizons}

%----------------------------------------------------------------------
%
%
%
%----------------------------------------------------------------------
%\section{Conclusion}

\section*{Acknowledgments}

RM thanks A. Ashtekar, A. \v Cade\v z, G,  Cook, P. Laguna, H.
Pfeiffer, and D. Shoemaker for insightful comments. 
%We especially thank
%P. Marronetti for extensive discussions on elliptic solver technique,
%and for a critique of an earlier version of this work. 
This work is
supported in part by NSF grants PHY~9800722, PHY~9800725, and
PHY~0102204. Computations were performed at the NSF supercomputer
center NCSA and the University of Texas AHPCC.

\end{document}